\documentclass[journal, twocolumn,twoside]{IEEEtran}

\normalsize

\usepackage{cite}

\ifCLASSINFOpdf
   \usepackage[pdftex]{graphicx}
\else
  \usepackage[dvips]{graphicx}
\fi

\usepackage[cmex10]{amsmath}
\usepackage{amssymb}

\usepackage{stfloats}

\usepackage{theorem} 

 \newtheorem{theorem}{Theorem}

  \newtheorem{definition}[theorem]{Definition} 
  \newtheorem{remark}[theorem]{Remark} 
\newcounter{mytempeqcounter}

\newcounter{example}
\newenvironment{example}[1][]{\refstepcounter{example}\par\medskip\noindent%
   \textbf{Example~\theexample. #1} \rmfamily}{\medskip}
   
   \newcounter{propertycounter}

\usepackage{color,soul}

\definecolor{lightblue}{rgb}{.90,.95,1}
\sethlcolor{yellow}

\begin{document}
%
\title{Spatially Coupled LDPC Codes\\
Constructed from Protographs}
%
%
%

\author{David~G.~M.~Mitchell,~\IEEEmembership{Member,~IEEE,}
Michael~Lentmaier,~\IEEEmembership{Senior Member,~IEEE,}
        and~Daniel~J.~Costello,~Jr.,~\IEEEmembership{Life~Fellow,~IEEE}
\thanks{This work was supported in part by the National Science Foundation under Grant Number CCF-$1161754$. The material in this paper was presented in part at the Information Theory and Applications Workshop, San Diego, CA, February 2009, Information Theory and Applications Workshop, San Diego, CA, February 2010, in part at the IEEE International Symposium on Information Theory, Austin, TX, July 2010, in part at the International Symposium on Turbo Coding and Iterative Information Processing, Brest, France, September 2010, in part at the Information Theory and Applications Workshop, San Diego, CA, February 2011, and in part at the IEEE International Symposium on Information Theory, St.~Petersburg, Russia, Aug. 2011.}
\thanks{D.~G.~M.~Mitchell and D.~J.~Costello,~Jr. are with the Department of
Electrical Engineering, University of Notre Dame, Notre Dame, IN 46556, USA
(e-mail: david.mitchell@nd.edu; costello.2@nd.edu).}
\thanks{M. Lentmaier is with the Department of Electrical and Information Technology, Lund University, Lund, Sweden (e-mail: Michael.Lentmaier@eit.lth.se).}
}

%
%

\markboth{IEEE Transactions on Information Theory (submitted paper)}%
{IEEE Transactions on Information Theory (submitted paper)}
%



\maketitle

\begin{abstract}
In this paper, we construct protograph-based spatially coupled low-density parity-check (SC-LDPC) codes by coupling together {a series of $L$ disjoint, or uncoupled, LDPC code Tanner graphs} into a single coupled chain. By varying $L$, we obtain a flexible family of code ensembles with varying rates and frame lengths that can share the same encoding and decoding architecture for arbitrary $L$. We demonstrate that the resulting codes combine the best features of optimized irregular and regular codes in one design: capacity approaching iterative belief propagation (BP) decoding thresholds \emph{and} linear growth of minimum distance with block length. In particular, we show that, for sufficiently large $L$, the BP thresholds on both the binary erasure channel (BEC) and the binary-input additive white Gaussian noise channel (AWGNC) \emph{saturate} to a particular value significantly better than the  BP decoding threshold and numerically indistinguishable from the optimal maximum a-posteriori (MAP) decoding threshold of the uncoupled LDPC code. When all variable nodes in the coupled chain have degree greater than two, asymptotically the error probability converges at least doubly exponentially with decoding iterations and we obtain sequences of asymptotically good LDPC codes with fast convergence rates and BP thresholds close to the Shannon limit. Further, the gap to capacity decreases as the density of the graph increases, opening up a new way to construct capacity achieving codes on memoryless binary-input symmetric-output (MBS) channels with low-complexity BP decoding.
\end{abstract}


%
\IEEEpeerreviewmaketitle

\section{Introduction}
%
%
%

%
\IEEEPARstart{T}{he} performance of an iterative belief propagation (BP) decoder for low-density parity-check (LDPC) codes is strongly influenced by the degrees of the different variable nodes and check nodes in the associated Tanner graph code representation \cite{tan81}. $(J,K)$-\emph{regular} LDPC codes, with constant variable node degree $J$ and check node degree $K$, as originally proposed by Gallager \cite{gal62} in 1962, are \emph{asymptotically good} in the sense that their minimum distance grows linearly with block length for $J>2$; however, the iterative decoding behavior of regular codes in the so-called \emph{waterfall}, or moderate bit error rate (BER), region of the performance curve falls short of capacity, making them unsuitable for severely power-constrained applications, such as uplink cellular data transmission or digital satellite broadcasting systems. 

On the other hand, optimized \emph{irregular} LDPC codes \cite{lmss01}, with a variety of different node degrees, exhibit capacity approaching performance in the waterfall but, unlike $(J,K)$-regular codes, are normally subject to an \emph{error floor}, a flattening of the BER curve that results in poor performance at high signal-to-noise ratios (SNRs), as a result of a non-vanishing fraction of degree two variable nodes, making them undesirable in applications such as data storage and optical communication that require very low decoded BERs. For {irregular} LDPC code ensembles, the degrees of the variable and check nodes are {often} modeled as random variables that are characterized by their \emph{degree distributions} $\lambda(x)$ and $\rho(x)$, respectively{ }\cite{lmss01}{.} Each coefficient in the polynomials $\lambda(x)$ and $\rho(x)$ corresponds to the fraction of edges in the graph connected to nodes of a certain degree.  {Gallager's $(J,K)$-regular LDPC code ensembles} correspond to the special case $\lambda(x)=x^{J-1}$ and $\rho(x)=x^{K-1}$, \emph{i.e.}, the degrees of each node type are constant. Using an algorithm called \emph{density evolution} (DE) \cite{ru01b}, a BP decoding threshold can be calculated for a randomly constructed LDPC code with degree distribution pair $(\lambda(x),\rho(x))$ that determines the limit of the error-free region asymptotically as the block length tends to infinity.  Using DE,  irregular code ensembles with thresholds very close to the Shannon limit on the binary-input additive white Gaussian noise channel (AWGNC) were designed in \cite{cfru01}. Moreover, in \cite{os02}, capacity achieving sequences of degree distribution pairs for a given rate $R$  with a vanishing gap between the threshold and the Shannon limit $\varepsilon_{Sh} = 1-R$ were presented for the binary erasure channel (BEC).  

LDPC convolutional codes (LDPC-CCs) \cite{fz99}, the convolutional counterparts of LDPC block codes (LDPC-BCs), have been shown to have certain advantages compared to LDPC-BCs \cite{cpbz06,cpjd07}. Variations in the check and variable node degrees of LDPC-CC Tanner graphs  are also characterized by a degree distribution pair, where the connections between nodes in the bi-infinite Tanner graph occur within a \emph{constraint length}. 
The performance of LDPC-CCs under iterative BP decoding has been well studied. Extensive computer simulation results (see, \emph{e.g.}, \cite{fz99,tss+04,pjs+08,psvc11}) have verified that, for practical code lengths, LDPC-CCs obtained by \emph{unwrapping} an LDPC-BC achieve a substantial \emph{convolutional gain} compared to the underlying LDPC-BC, where both codes have the same computational complexity with iterative decoding and the block length of the LDPC-BC equals the constraint length of the LDPC-CC. Moreover, various code and graph properties, such as iterative decoding thresholds \cite{slcz04,lszc05,lscz10}, girth \cite{ltz01,psvc11}, minimum (free) distance \cite{tss+04,stl+07,tzc10,mpc13}, minimum (free) pseudo-distance  \cite{spvc09}, and minimum trapping set size \cite{mpc13}, of the unwrapped LDPC-CC have been shown to be at least as good as the corresponding values of the underlying LDPC-BC. 

Spatially coupled LDPC (SC-LDPC) codes are constructed by coupling together {a series of $L$ disjoint, or uncoupled, LDPC code Tanner graphs} into a single coupled \emph{chain}. They can be viewed as a type of LDPC-CC, since spatial coupling is equivalent to introducing memory into the encoding process. If the coupled chain is \emph{unterminated} ($L\rightarrow \infty$), a SC-LDPC convolutional code (SC-LDPC-CC) is formed, and if the chain is terminated (finite $L$), a SC-LDPC block code (SC-LDPC-BC) results. Recently, it has been proven by Kudekar \emph{et al.} that SC-LDPC-BC ensembles are capacity achieving on memoryless binary-input symmetric-output (MBS) channels under BP decoding \cite{kru11,kru13}. Consequently, the principle of spatial graph coupling has attracted significant attention and has been successfully applied in many other areas of communications and signal processing, such as, for example, compressed sensing \cite{kp10,djm13}, relay channels \cite{sts11,uks11,sgm12,wcf13}, wiretap channels \cite{ruas11}, multiple access channels \cite{kk11, ynpn11,tru12b,st13}, broadcast channels \cite{stso12}, intersymbol-interference channels \cite{kk11b,nypn12}, multiuser detection \cite{ttk11}, random access \cite{lplc12}, source coding \cite{amuv12}, quantum codes \cite{hkis11,amt12},  and models in statistical physics \cite{hmu10}. Also, studies of the finite length scaling properties of SC-LDPC-BCs were performed in \cite{ou11,ou13} and block erasure channel performance bounds were given in \cite{ja13}. 

LDPC code ensembles with a certain predefined \emph{structure} can be constructed by means of \emph{protographs} \cite{tho03}. By applying a graph lifting operation, Tanner graphs of various sizes can be constructed that preserve the rate, degree distribution, and \emph{computation graphs} (see \cite{ru08}) of the protograph. It has been observed that irregular protograph-based LDPC-BC ensembles often have better thresholds than \emph{unstructured} irregular ensembles with the same degree distributions \cite{ddja09}. {An extreme example of this behavior is that the thresholds of carefully designed protograph-based LDPC code ensembles containing variable nodes of degree one can have good thresholds}\cite{ddja09}{; whereas an unstructured LDPC code ensemble with degree one variable nodes will not even have a threshold.} Moreover, the inherent structure in protograph-based ensembles can improve distance properties. For example, irregular protograph-based LDPC-BC ensembles that contain degree two variable nodes can be asymptotically good, and ensembles with minimum variable node degree three can provide a good trade-off between distance and threshold \cite{ddja09}. As a result of their good properties and implementation advantages, many LDPC codes have been adopted in recent industry standards, such as wireless LANs (IEEE 802.11n), WiMax (IEEE 802.16e), digital video broadcasting (DVB-S2), 
and the ITU-T standard for networking over power lines, phone lines, and coaxial cable (G.hn/G.9960), and each of these standard codes can be viewed as protograph-based LDPC-BCs.


In this paper, we analyze ensembles of SC-LDPC-BCs constructed from protographs. We present an \emph{edge spreading} procedure that is used to couple together $L$ block protographs to form a convolutional protograph. The protograph framework enables us to extend previous DE analysis \cite{slcz04,lszc05,lscz10} and codeword weight enumerator analysis \cite{stl+07} that were restricted to certain $(J,K)$-regular SC-LDPC-CC ensembles to general $(J,K)$-regular and irregular ensembles. We use this analysis to show that, for protograph-based SC-LDPC-BC ensembles with sufficiently large $L$, the iterative BP decoding thresholds on both the BEC and the AWGNC \emph{saturate} to a particular value significantly larger than the BP decoding threshold and numerically indistinguishable from the maximum a-posteriori (MAP) decoding threshold of the underlying LDPC-BC ensemble. Further, for $(J,K)$-regular ensembles with $J>2$ and properly designed irregular ensembles, we show that SC-LDPC-BCs are asymptotically good, \emph{i.e.}, their minimum distance grows linearly with block length. Thus, since the MAP thresholds of $(J,K)$-regular LDPC-BC ensembles approach capacity as the graph density increases, protograph-based SC-LDPC-BC ensembles combine the best features of optimized irregular and regular codes in one design: capacity approaching BP decoding thresholds \emph{and} linear minimum distance growth. Finally, we study the relationship between the minimum distance growth rate of the SC-LDPC-BC ensemble and the free distance growth rate of the associated SC-LDPC-CC ensemble.

The paper is structured as follows. In Section \ref{sec:ldpccc}, we give a brief review of LDPC-BCs and the protograph construction method. We then describe the construction and structural properties of protograph-based SC-LDPC-CC and SC-LDPC-BC ensembles. In Section  \ref{sec:tradeoff}, we begin with an asymptotic weight enumerator analysis of protograph-based SC-LDPC-BC ensembles. We then proceed to study their iterative decoding properties by means of DE, first for the BEC and then for the AWGNC. As the coupling length $L$ increases, we obtain a family of asymptotically good code ensembles with increasing rates that feature a trade-off between capacity approaching iterative decoding thresholds and declining minimum distance growth rates. Then, in Section \ref{sec:growth}, we show that the minimum distance growth rates, while declining with $L$, converge to a bound on the free distance growth rate of the unterminated SC-LDPC-CC ensemble that is independent of $L$ and significantly larger than the minimum distance growth rate of the underlying LDPC-BC ensemble. We then argue that an appropriate distance measure for terminated  SC-LDPC-CC (\emph{i.e.}, SC-LDPC-BC) ensembles should also behave independently of $L$. Some concluding remarks are given in Section \ref{sec:conc}.

\section{Spatially Coupled LDPC Code Ensembles}\label{sec:ldpccc}
In this section, we will describe the construction of protograph-based SC-LDPC code ensembles. We begin with a brief introduction to LDPC codes in Section~\ref{sec:ldpc} and review the construction of LDPC code ensembles based on protographs in Section~\ref{sec:proto}. In Section~\ref{sec:convproto} we discuss the construction of a \emph{convolutional protograph} and the associated ensemble of protograph-based SC-LDPC-CCs by applying an edge spreading operation to (spatially) couple together a sequence of uncoupled LDPC-BC protographs. In Section~\ref{sec:convtoblock}, we present two closely related ways to construct SC-LDPC-BCs from protograph-based SC-LDPC-CCs: termination and tail-biting. We conclude with a discussion of variations to the edge spreading rule and different ways of constructing SC-LDPC-BC ensembles in Section~\ref{sec:scdiscuss}.

\subsection{LDPC Block Codes}\label{sec:ldpc}

We begin with a brief introduction to LDPC-BCs (see also \cite{lc04,ru08}). A $(J,K)$\emph{-regular} LDPC-BC  is defined as the null space of a \emph{sparse} binary parity-check matrix $\mathbf{H}$, where each row of $\mathbf{H}$ contains exactly $K$ ones, each column of $\mathbf{H}$ contains exactly $J$ ones, and both $J$ and $K$ are small compared with the number of rows in $\mathbf{H}$. An LDPC-BC code is called \emph{irregular} if the row and column weights are not constant. The code has \emph{block length} $n$, where $n$ is the number of columns of $\mathbf{H}$, and \emph{rate} $R=k/n$, where $(n-k)$ is the rank of $\mathbf{H}$. For $(J,K)$-regular codes, the code rate is given as $R\geq1-J/K$, with equality when $\mathbf{H}$ has full rank. It is often useful to represent the parity-check matrix $\mathbf{H}$ using a bipartite graph called the \emph{Tanner graph} \cite{tan81}. In the Tanner graph representation, each column of $\mathbf{H}$ corresponds to a \emph{code bit} or \emph{variable node} and each row corresponds to a \emph{parity-check} or \emph{check node}. If position $(i,j)$ of $\mathbf{H}$ is equal to one, then check node $i$ is connected by an \emph{edge} to variable node $j$ in the Tanner graph; otherwise, there is no edge connecting these nodes. The notion of \emph{degree distribution} is used to characterize the variations of check and variable node degrees  (see \cite{lmss01}).

\subsection{Protograph-based Code Construction}\label{sec:proto}
\begin{figure}[t]
\begin{center}
\includegraphics{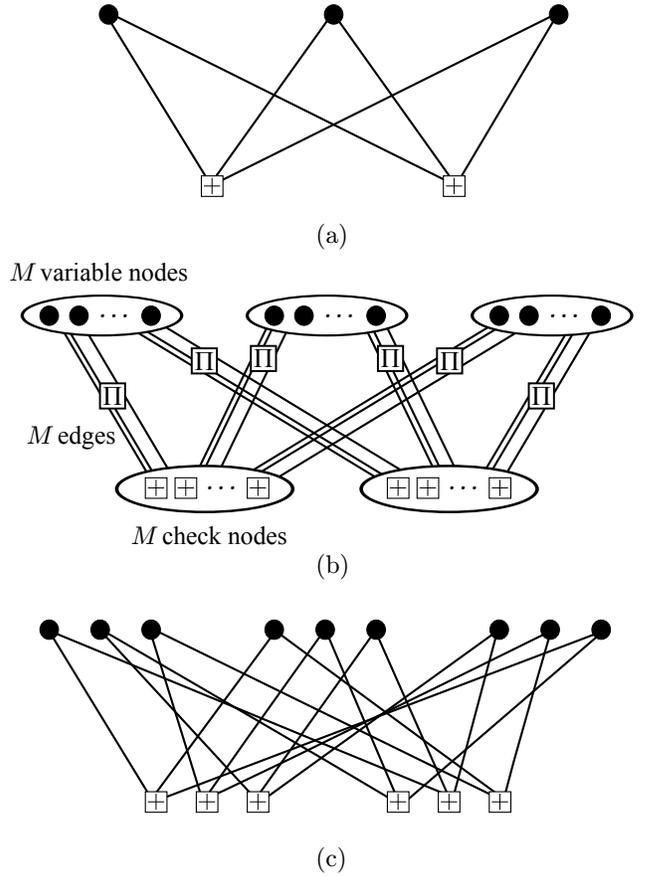} 
\end{center}
\caption{(a) Protograph of a $(2,3)$-regular LDPC-BC ensemble, (b) Tanner graph of a $(2,3)$-regular LDPC-BC lifted from the protograph with lifting factor $M$, where $\Pi$ denotes a random interleaver (permutation) of size $M$, and (c) example $(2,3)$-regular Tanner graph lifted from the protograph with $M=3$.}\label{fig:lift}
\end{figure}

A \emph{protograph} \cite{tho03} with \emph{design rate} $R=1-n_c/n_v$ is a small bipartite graph $(V,C,E)$ that connects a set of $n_v$ variable nodes $V=\{v_0,\ldots,v_{n_v-1}\}$ to a set of $n_c$ check nodes $C=\{c_0,\ldots,c_{n_c-1}\}$ by a set of edges $E$. We assume $n_v>n_c$ so that the protograph has a strictly positive design rate. Fig.~\ref{fig:lift}(a) shows an example protograph with $n_v = 3$ variable nodes and $n_c=2$ check nodes. The Tanner graph representing a \emph{protograph-based} LDPC-BC with block length $n=Mn_v$ is obtained by taking an $M$\emph{-fold graph cover} (see \cite{psvc11}) or ``$M$\emph{-lifting}'' of the protograph.  Graph lifting can be informally described as follows: each edge in the protograph becomes a bundle of $M$ edges, connecting $M$ copies of a variable node to $M$ copies of a check node. The connections within each bundle are then permuted between the variable and check node pairings. The resulting covering graph is $M$ times larger than the protograph and has the same rate, degree distribution, and computation graphs as the protograph.\footnote{The computation graphs are preserved since the computation graph of each of the $M$ copies of a variable node in the lifted graph is identical to the computation graph of the original variable node in the protograph (see \cite{tho03}).} 

\begin{example}\label{ex:23proto}
Fig.~\ref{fig:lift}(b) shows a general $M$-lifting of the $(2,3)$-regular protograph given in Fig.~\ref{fig:lift}(a), and Fig.~\ref{fig:lift}(c) shows a particular $M$-lifting with $M=3$. \end{example}\hfill $\Box$

The protograph can be represented by its $n_c\times n_v$ \emph{base} biadjacency matrix $\mathbf{B}$, where $B_{i,j}$ is taken to be the number of edges connecting variable node $v_j$ to check node $c_i$. In general, a protograph can have multiple edges connecting a variable node to a check node, which corresponds to entries in $\mathbf{B}$ greater than $1$. The $Mn_c \times Mn_v$ parity-check matrix $\mathbf{H}$ of a  protograph-based LDPC-BC  with block length $n=Mn_v$ and design rate $R=1-Mn_c/Mn_v=1-n_c/n_v$ is created ($M$-\emph{lifted}) by replacing each non-zero entry in $\mathbf{B}$  by a sum of $ B_{i,j}$  permutation matrices of size $M\times M$ and each zero entry by the $M\times M$ all-zero matrix.\footnote{The lifted parity-check matrix $\mathbf{H}$ may have linearly dependent rows; this simply means that the lifted code has a slightly higher rate than the design rate $R=1-n_c/n_v$.} 

\par\medskip\noindent{\bf Example \ref{ex:23proto} (cont.)} The base matrix of the protograph shown in Fig.~\ref{fig:lift}(a) is
\begin{equation}\label{23base}\mathbf{B}=\left[
\begin{array}{ccccc}
1 & 1 & 1\\
1 & 1 & 1
\end{array}\right],
\end{equation}
\noindent 
and the parity-check matrix corresponding to an $M$-lifting of the base matrix in (\ref{23base}) is
\begin{equation}\mathbf{H}=\left[
\begin{array}{ccccc}
\mathbf{\Pi}_{1,1} & \mathbf{\Pi}_{1,2} & \mathbf{\Pi}_{1,3}\\
\mathbf{\Pi}_{2,1} & \mathbf{\Pi}_{2,2} & \mathbf{\Pi}_{2,3}
\end{array}\right],\end{equation}
where  $\mathbf{\Pi}_{i,j}$ is an $M\times M$ permutation matrix. \hfill $\Box$\medskip

Since an LDPC-BC is defined as the null space of a sparse parity-check matrix $\mathbf{H}$, we define the \emph{ensemble} of protograph-based LDPC-BCs with block length $n = M n_v$ and design rate $R=1-n_c/n_v$ as the set of all parity-check matrices $\mathbf{H}$ that can be lifted from a given base matrix $\mathbf{B}$, or equivalently as the collection of all $M$-fold cover graphs of the protograph. It is an important feature of this construction that each lifted code inherits the design rate, degree distribution, and computation graphs of the protograph.  As a consequence, {ensemble} DE and weight enumerator analysis can be performed \emph{within} the protograph \cite{ddja09}. Using these tools, properly designed protograph-based LDPC-BC ensembles have been shown in the literature to have many desirable features, such as good iterative decoding thresholds and linear minimum distance growth (see, e.g., \cite{ddja09}). 

\begin{figure}[t]
\begin{center}
\includegraphics{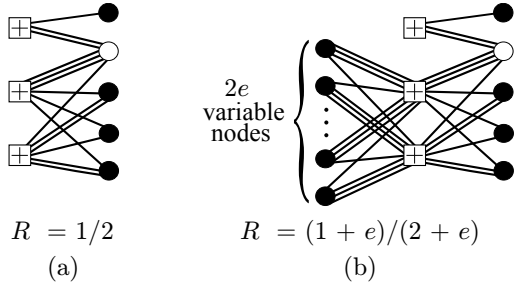}\end{center}
\caption{Protographs representing good irregular LDPC-BC ensembles: (a) the ARJA protograph with design rate $R=1/2$, and (b) the family of AR4JA protographs with extension parameter $e$ and design rates $R=(1+e)/(2+e)$. White circles represent punctured variable nodes.}\label{fig:ar4ja}
\end{figure}

A particularly interesting example of such a code design that incorporates both of these desirable features is the accumulate-repeat-jagged-accumulate (ARJA) and accumulate-repeat-by-$4$-jagged-accumulate (AR4JA) family of irregular protograph-based LDPC-BC ensembles \cite{ddja09}. These practically interesting codes have recently been selected as a CCSDS standard for near-earth and deep space communication \cite{ccsds07} and serve in this paper as an example and template for the successful application of spatial coupling to irregular graphs. The protographs of these ensembles are depicted in Fig.~\ref{fig:ar4ja}. (Note that setting $e=0$ in the AR4JA protograph results in the ARJA protograph.) The white circles in these protographs represent \emph{punctured} variable nodes, \emph{i.e}, no code bits are transmitted in these positions. In a Tanner graph $M$-lifted from the protograph, the $M$ copies of a punctured variable node are also punctured. The design rate of a protograph-based LDPC-BC ensemble with $n_t$ transmitted variable nodes in the protograph is
\begin{equation}\label{puncrate}
R=\frac{n_v-n_c}{n_t}.
\end{equation}
Note that, in the case $n_t=n_v$, there is no puncturing and consequently the design rate is $R=1-n_c/n_v$.

\subsection{Convolutional Protographs and Spatial Coupling}\label{sec:convproto}
In this section, we introduce the notion of a \emph{convolutional protograph}, which represents an ensemble of SC-LDPC-CCs. A convolutional protograph is obtained by connecting, or \emph{spatially coupling}, a sequence of disjoint block protographs together in a chain. Spatial coupling introduces \emph{memory} into the code design, \emph{i.e.}, transitioning from a block code to a convolutional code, and be achieved by applying an \emph{edge spreading} operation to the sequence of disjoint block protographs.

\medskip\noindent{\bf Edge Spreading Rule for Spatial Coupling:} Consider replicating a block protograph with $b_v$ variable nodes and $b_c$ check nodes as an infinite sequence of disjoint graphs. We associate each graph in the sequence with a time index $t$. Suppose variable node $v_j$ is connected to check node $c_i$ by $B_{i,j}$ edges in each protograph, where $i\in\{0,1,\ldots,b_c-1\}$ and $j\in\{0,1,\ldots,b_v-1\}$. We now \emph{spread} (connect) the $B_{i,j}$ edges emanating from node $v_j$ at time $t$ arbitrarily over the $w+1$ check nodes of type $c_i$ at times $t,t+1,\ldots,t+w$, where $w>0$ is the \emph{coupling width} of the graph, or \emph{memory} of the code.\footnote{The coupling width $w$ is referred to in convolutional coding parlance as the \emph{syndrome former memory} (see, \emph{e.g.} \cite{fz99}), or, in the recent series of papers by Kudekar \emph{et al.}, the  \emph{smoothing parameter} \cite{kru11,kru13}.} This operation is repeated (independently) for each of the $b_v$ variable nodes at time $t$. Applying this edge spreading identically to the variable nodes at all time instants results in a \emph{convolutional protograph}.\medskip



\begin{definition}\label{def:ldpccc}\vspace{-2mm}
\emph{An ensemble of protograph-based spatially coupled LDPC-CCs (SC-LDPC-CCs) with coupling width $w$, design rate $R=1-b_c/b_v$, and \emph{constraint length} $\nu=(w+1)Mb_v$  is the collection of all $M$-fold graph covers of a convolutional protograph.}
\end{definition}

In this paper we are primarily interested in asymptotic results in the code block or constraint length, \emph{i.e.}, in the regime where the lifting factor $M$ tends to infinity. A block/convolutional protograph represents a finite block/constraint length LDPC-BC/LDPC-CC ensemble for each lifting factor $M$; however, in the sequel, unless stated otherwise, we consider an infinite lifting factor. Thus, if we refer to a code ensemble represented by a protograph in the singular, we implicitly assume infinite $M$. 

Note that the convolutional protograph constructed using the Edge Spreading Rule has the same design rate, degree distribution, and computation graphs as the original block protograph. This ensures that the computation graphs of the LDPC-CC ensembles defined by the convolutional protograph are the same as those of the LDPC-BC ensembles defined by the original block protograph. There are many ways to apply the Edge Spreading Rule for a given $w$ to a sequence of disjoint block protographs that give an extra degree of freedom in the protograph-based construction. It will be shown later that different edge spreadings affect both the iterative BP decoding performance and distance properties of the resulting code ensemble. Generalizations of the edge spreading operation and a discussion of the degrees of freedom in the design are presented in Section \ref{sec:scdiscuss}.

\begin{figure}[t]
\begin{center}
\includegraphics{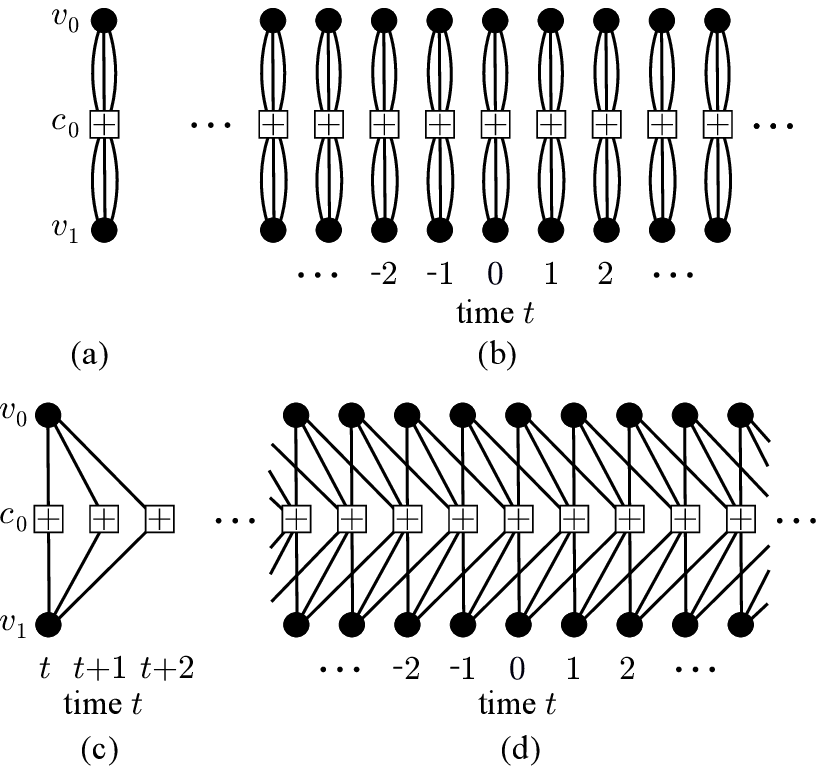} 
\end{center}
\caption{(a) Protograph representing a $(3,6)$-regular LDPC-BC ensemble, (b) sequence of $(3,6)$-regular LDPC-BC protographs, (c) illustration of edge spreading for one segment of the graph at time $t$ with coupling width $w=2$, and (d) protograph representing a (spatially coupled) $(3,6)$-regular LDPC-CC ensemble with $w=2$.}\label{fig:edgespreading}
\end{figure}

\begin{example}\label{ex:36proto} Fig.~\ref{fig:edgespreading} illustrates the edge spreading operation applied to a $(3,6)$-regular (block) protograph with design rate $R=1/2$. First, the protograph is replicated as an infinite sequence of disjoint graphs, shown in Fig.~\ref{fig:edgespreading}(b). (This can be considered as block code transmission over time.) An edge spreading with coupling width $w=2$ applied to the variable nodes at time $t$ is shown in Fig.~\ref{fig:edgespreading}(c). The three edges emanating from each variable node $v_0$ and $v_1$ are spread such that exactly one edge connects the variable node to check node $c_0$ at times $t$, $t+1$, and $t+2$.  Applying this edge spreading to variable nodes at all time instants results in the $(3,6)$-regular convolutional protograph with design rate $R=1/2$ shown in Fig.~\ref{fig:edgespreading}(d). \hfill $\Box$
\end{example}

The (convolutional) base matrix corresponding to the convolutional protograph is
\begin{equation}\label{convbase}\mathbf{B}_{[-\infty,\infty]}=\left[
\begin{array}{lllllll}
&\multicolumn{1}{l}{\hspace{-2mm}\ddots} &  \multicolumn{1}{l}{\hspace{-2mm}\ddots}& \multicolumn{2}{c}{\hspace{3mm}\ddots} \\ 
&\mathbf{B}_{w} &\mathbf{B}_{w-1}& \cdots  & \mathbf{B}_0 & &\\
&&\mathbf{B}_{w} &\mathbf{B}_{w-1}& \cdots  & \mathbf{B}_0 & \\
& &\multicolumn{1}{r}{\ddots} &  \multicolumn{1}{r}{\ddots}   & &\multicolumn{2}{r}{\ddots}  \\
\end{array}\right],\end{equation}
where the $b_c \times b_v$ {\em component base matrices} $\mathbf{B}_{i}$, $i=0,1,\dots,w$, represent the edge connections from the $b_v$ variable nodes at time $t$ to the $b_c$ check nodes at time $t+i$. Starting from the base matrix $\mathbf{B}=[B_{i,j}]_{0\leq i\leq n_c-1,0\leq j\leq n_v-1}$ of an LDPC-BC ensemble, the Edge Spreading Rule divides the edges associated with each variable node in $\mathbf{B}$ among $w+1$ component base matrices $\mathbf{B}_i$, $i=0,1,\dots,w$,   such that the condition 
\begin{equation}\label{edgespread}\sum_{i=0}^w \mathbf{B}_i=\mathbf{B}\end{equation}
 is satisfied, where each $\mathbf{B}_i$ contains non-negative integer entries.

\par\medskip\noindent{\bf Example \ref{ex:36proto} (cont.).} The $(3,6)$-regular protograph shown in Fig.~\ref{fig:edgespreading}(a) has base matrix $\mathbf{B} = [\>
3 \>\>\>\> 3\>]$. The edge spreading depicted in Fig.~\ref{fig:edgespreading}(c) with $w=2$ corresponds to component base matrices
\begin{equation}\label{eq:e
x36}
\mathbf{B}_0=[\>
1 \>\>\>\> 1\>]=\mathbf{B}_1=\mathbf{B}_2.
\end{equation}
(Note that this is a valid edge spreading since the component base matrices conform to condition (\ref{edgespread}).) Then the $(3,6)$-regular convolutional base matrix corresponding to the convolutional protograph of Fig.~\ref{fig:edgespreading}(d) is obtained in the form of (\ref{convbase}) as
$$
 \mathbf{B}_{[-\infty,\infty]}=\left[
\begin{array}{cccccccccc}
  \ddots\hspace{-2mm} &   & \ddots\hspace{-2mm} &   & \ddots\hspace{-1.8mm} & & & & &\\\cline{1-6}
\multicolumn{1}{|c}{1} & 1 & \multicolumn{1}{|c}{1} & 1 & \multicolumn{1}{|c}{1} & \multicolumn{1}{c|}{1} & & & &\\\cline{1-8}
 & & \multicolumn{1}{|c}{1} & 1 & \multicolumn{1}{|c}{1} & 1 & \multicolumn{1}{|c}{1} & \multicolumn{1}{c|}{1}  & &\\\cline{3-10}
& & & & \multicolumn{1}{|c}{1} & 1 & \multicolumn{1}{|c}{1} & 1 & \multicolumn{1}{|c}{1} & \multicolumn{1}{c|}{1} \\\cline{5-10}
& & & & & \ddots\hspace{-2mm} & & \ddots\hspace{-2mm} & & \ddots\hspace{-2mm}
\end{array}
\right].
$$ \hfill $\Box$\medskip

A similar edge spreading to that used in Example \ref{ex:36proto} can be applied to construct $(J,K)$-regular convolutional protographs from $(J,K)$-regular block protographs with $\textrm{gcd}(J,K) >1$.

\medskip\noindent{\bf The $\mathcal{C}(J,K)$ SC-LDPC-CC Ensemble:} Let $a=\textrm{gcd}(J,K)$ denote the greatest common divisor of $J$ and $K$. Then there exist positive integers $J^\prime$ and $K^\prime$ such that $J=aJ^\prime$ and $K=aK^\prime$ with $\textrm{gcd}(J^\prime,K^\prime)=1$. It follows that the base matrix of a $(J,K)$-regular protograph-based SC-LDPC-CC ensemble with coupling width $w=a-1$ can be defined as in (\ref{convbase}), where the submatrices $\mathbf{B}_i$, $i=0,\ldots,w$, are identical $J^\prime\times K^\prime$ matrices with all entries equal to one.  We denote the SC-LDPC-CC ensembles constructed using this edge spreading as $\mathcal{C}(J,K)$.\medskip

\noindent Note that, if $a=1$, the coupling width is equal to zero and the convolutional protograph is not fully connected. In this case, we can simply choose a different edge spreading of a $(J,K)$-regular block protograph following the Edge Spreading Rule. 
\begin{example}\label{ex:34} Consider the $(3,4)$-regular protograph defined by the all-ones base matrix $\mathbf{B}$ of size $3\times 4$. We can spread the edges of $\mathbf{B}$ as
\begin{equation*}\label{34component}
\mathbf{B}_0=\left[\begin{array}{ccccc}
1 & 1 & 0 & 0\\
0& 1 & 1 & 0\\
0 & 0 & 1 & 1\\
\end{array}\right]\textrm{ and }\mathbf{B}_1=\left[\begin{array}{ccccc}
0 & 0 & 1 & 1\\
1 & 0 & 0 & 1\\
1 & 1 & 0 & 0\\
\end{array}\right].
\end{equation*}
These component base matrices satisfy condition (\ref{edgespread}) and can be used to construct a $(3,4)$-regular convolutional base matrix $\mathbf{B}_{[-\infty,\infty]}$ with coupling width $w=1$ (see \eqref{convbase}). \end{example}\hfill $\Box$

 Next, we demonstrate the application of the Edge Spreading Rule to irregular protographs.

\begin{figure}[t]
\begin{center}
\includegraphics{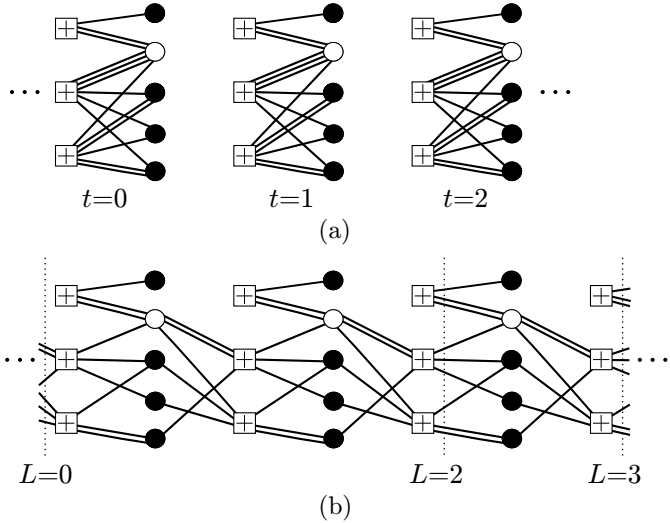} 
\end{center}
\caption{(a) Sequence of ARJA block protographs with design rate $R=1/2$, and (b)  spatially coupled ARJA convolutional protograph with $w=1$ and design rate $R=1/2$. Also shown are the termination markings for the related spatially coupled ARJA block protograph.}\label{fig:arja}
\end{figure}

\begin{figure}[t]
\begin{center}
\includegraphics{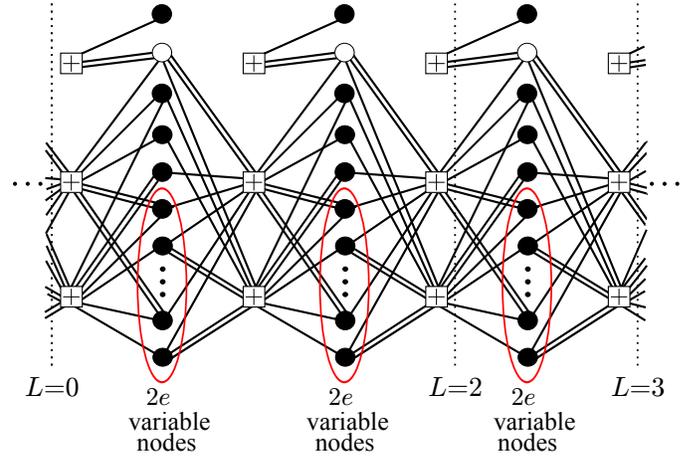} 
\end{center}
\caption{Spatially coupled AR4JA convolutional protographs with coupling width $w=1$ and design rate $R=(1+e)/(2+e)$. Also shown are the termination markings for the related spatially coupled AR4JA block protograph.}\label{fig:ar4jacc}
\end{figure}

\begin{example}\label{ex:arja} Fig.~\ref{fig:arja} shows an example of the Edge Spreading Rule applied to the irregular ARJA protograph with base matrix \begin{equation}\label{arjacomponent}
\mathbf{B}=\left[\begin{array}{ccccc}
1 & 2 & 0 & 0 & 0\\
0 & 3 & 1 & 1 & 1\\
0 & 1 & 2 & 1 & 2\\
\end{array}\right].
\end{equation} A sequence of disjoint ARJA protographs with design rate $R=1/2$ is shown in Fig.~\ref{fig:arja}(a). An irregular (spatially coupled) ARJA convolutional protograph with design rate $R=1/2$ and coupling width $w=1$ is shown in Fig.~\ref{fig:arja}(b). The  component base matrices corresponding to this edge spreading are
\begin{equation*}\label{arjacomponent}
\mathbf{B}_0=\left[\begin{array}{ccccc}
1 & 2 & 0 & 0 & 0\\
0 & 1 & 1 & 1 & 0\\
0 & 0 & 1 & 0 & 2\\
\end{array}\right], \mathbf{B}_1=\left[\begin{array}{ccccc}
0 & 0 & 0 & 0 & 0\\
0 & 2 & 0 & 0 & 1\\
0 & 1 & 1 & 1 & 0\\
\end{array}\right],
\end{equation*}
where $\mathbf{B}_0+\mathbf{B}_1=\mathbf{B}$. Note that there is one punctured variable node at each time instant of the convolutional protograph. In the sequel, we refer to the SC-LDPC-CC ensemble represented by this ARJA convolutional protograph by $\mathcal{C}_{\mathrm{ARJA}}$.

Similarly, we can construct a series of AR4JA convolutional protographs with coupling width $w=1$ and design rate $R=(1+e)/(2+e)$ using the edge spreading shown in Fig.~\ref{fig:ar4jacc}. The resulting SC-LDPC-CC ensembles are denoted $\mathcal{C}_{\mathrm{AR4JA}}(e)$.
\end{example}\hfill $\Box$


\begin{remark} \emph{The convolutional protograph constructed using the Edge Spreading Rule with coupling width $w$ can be viewed as an infinite graph lifting of the block protograph. Consequently, a protograph-based SC-LDPC-CC can be viewed as a double graph cover of a block code protograph. As the local connectivity is maintained by graph lifting, the computation graph is identical and the BP decoder cannot distinguish if it is operating on the original protograph or a covering graph of the protograph. As a result, the BP decoding threshold of the SC-LDPC-CC is \emph{equal} to the BP decoding threshold of the uncoupled LDPC-BC ensemble. For further discussion of SC-LDPC-CCs as graph covers of LDPC-BCs, see \cite{tss+04,psvc11}}\end{remark}



\subsection{Spatially Coupled LDPC Block Codes}\label{sec:convtoblock}

Even though the convolutional protograph and lifted SC-LDPC-CC Tanner graphs extend infinitely forward and backward in time, in practice there is always some finite starting and ending time, \emph{i.e.}, the protograph is \emph{terminated}. As a consequence, `convolutional-like' block codes of flexible frame length can be obtained by termination, and we will see later that the iterative BP threshold of SC-LDPC-CCs is significantly improved by termination.. 

\subsubsection{Terminated SC-LDPC-CC ensembles} \label{sec:terminated}

\begin{definition}\label{def:termldpccc}\vspace{-2mm}
\emph{A terminated convolutional protograph with coupling width $w$ and \emph{coupling length} $L>0$ can be obtained as the subgraph of the convolutional protograph induced by the variable nodes over time instants $t=0,1,\ldots,L-1$. An ensemble of protograph-based \emph{spatially coupled LDPC-BCs (SC-LDPC-BCs)} with coupling width $w$, coupling length $L$, and block length $n=Mn_v=MLb_v$ is obtained as the collection of all $M$-fold graph covers of the terminated convolutional protograph, where $n_v=Lb_v$ is the total number of variable nodes in the terminated convolutional protograph.}
\end{definition}



Terminating the convolutional protograph is equivalent to applying the Edge Spreading Rule to spatially couple $L$ disjoint copies of a block protograph, where connections are allowed at the right hand boundary to $wb_c$ additional check nodes in sections $t=L, L+1, \ldots, L+w-1$. Consequently, there are $b_v$ variable nodes and $b_c$ check nodes at each time instant $t=0,1, \ldots, L-1$ and $b_c$ additional check nodes at each time instant $t=L, L+1, \ldots, L+w-1$. We now use $n_v$ and $n_c$ to denote the total number of variable nodes and check nodes, respectively, in the terminated convolutional protograph. The $n_c\times n_v =(L+w)b_c \times L b_v$ base matrix $\mathbf{B}_{[0,L-1]}$ corresponding to the terminated convolutional protograph is
\begin{equation}\label{termbase}\mathbf{B}_{[0,L-1]}=\left[
\begin{array}{cccccc}
\mathbf{B}_0\vspace{1mm} & &&&\\
\mathbf{B}_1 &\mathbf{B}_0&& &\\
\vdots & \mathbf{B}_1&&\ddots & & \\
\mathbf{B}_{w} & \vdots &&\ddots &&\mathbf{B}_0\vspace{1mm}  \\
&\mathbf{B}_{w}  & &&&\mathbf{B}_1 \\
&&& \ddots&& \vdots\\
&& &&&  \mathbf{B}_{w}\\
\end{array}\right]_{(L+w)b_c \times L b_v},\end{equation}
which can be obtained by truncating the convolutional base matrix $\mathbf{B}_{[-\infty,\infty]}$ from (\ref{convbase}).

Terminating a convolutional protograph has two effects on the resulting subgraph:
\begin{itemize}
\item A \emph{structured irregularity} is introduced to the graph: the variable nodes at each time index have the same number and type of edge connections as the original block protograph; however, the first $wb_c$ and last $wb_c$ check nodes have, in general, a reduced number of edge connections; 
\item For finite $L$, a \emph{rate loss} is incurred due to the check nodes at the right hand boundary of the subgraph (connections to check nodes at time instants $t=L,L+1,\ldots,L+w-1$).
\end{itemize}

\noindent The design rate of the terminated convolutional protograph (without puncturing) is
\begin{equation}\label{termrate}R_L=1-\frac{n_c}{n_v}=1-\frac{(L+w)b_c}{Lb_v}=1-\left(\frac{L+w}{L}\right)\left(1-R\right),\end{equation}
where $R=1-b_c/b_v$ is the design rate of the unterminated convolutional protograph (and the uncoupled block protograph). We assume that a sufficiently large $L$ is chosen such that the SC-LDPC-BC ensemble has a strictly positive design rate, \emph{i.e.}, $L>wb_c/(b_v-b_c)$. Assuming $w>0$ and finite $L$, we see from (\ref{termrate}) that the terminated convolutional protograph has reduced design rate $R_L < R$. Note that the rate loss and structured irregularity in the check node degree distribution introduced by termination become vanishingly small as the coupling length $L$ increases: the rate $R_L$ increases monotonically and approaches the design rate $R$ of the unterminated convolutional protograph ($\lim_{L\rightarrow\infty}R_L=R$), and the check node degree distribution approaches that of the unterminated convolutional protograph.

\begin{figure}[t]
\begin{center}
\includegraphics{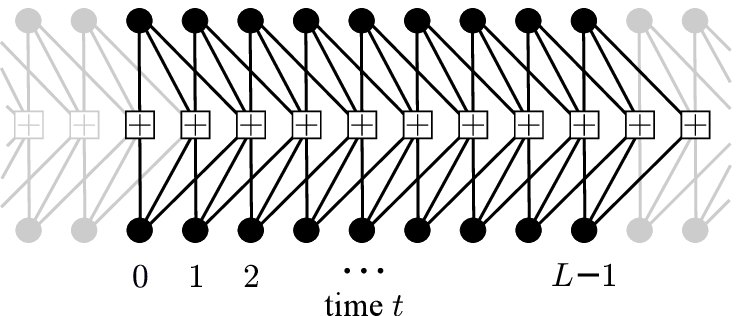} 
\end{center}
\caption{Protograph of a SC-LDPC-BC ensemble (highlighted in black) with coupling length $L$ and coupling width $w=2$ obtained by terminating a $(3,6)$-regular convolutional protograph.}\label{fig:term36}
\end{figure}

{\bf Example \ref{ex:36proto} (cont.)} Fig.~\ref{fig:term36} shows (highlighted in black) the terminated $(3,6)$-regular convolutional protograph induced by the variable nodes over time instants $t=0,1,\ldots,L-1$.  The corresponding terminated convolutional base matrix, obtained using (\ref{termbase}), is
$$\mathbf{B}_{[0,L-1]}=\left[\begin{array}{ccccccc}\cline{1-2}
\multicolumn{1}{|c}{1} & \multicolumn{1}{c|}{1}  &   &   \\
 \cline{1-4}                         
  \multicolumn{1}{|c}{1}    & \multicolumn{1}{c|}{1} & 1 & \multicolumn{1}{c|}{1} &           &    \\\cline{1-4}                                  
\multicolumn{1}{|c}{1} & \multicolumn{1}{c|}{1} & 1 & \multicolumn{1}{c|}{1} & \raisebox{-1.8mm}{$\ddots$} &\\ \cline{1-4}                                 
   &   & \multicolumn{1}{|c}{1} & \multicolumn{1}{c|}{1} & \raisebox{-1.8mm}{$\ddots$} &\\\cline{3-4}\cline{6-7}                                       
   &    &    &    & \raisebox{-1.8mm}{$\ddots$} & \multicolumn{1}{|c}{1} & \multicolumn{1}{c|}{1}\\\cline{6-7}
    &    &    &    &            & \multicolumn{1}{|c}{1} & \multicolumn{1}{c|}{1} \\\cline{6-7} 
    &    &    &    &            & \multicolumn{1}{|c}{1} & \multicolumn{1}{c|}{1} \\\cline{6-7} 
\end{array}\right]_{(L+2)\times 2L}.$$

Note the structured irregularity in the resulting chain of coupled protographs: each variable node is connected to $3$ check nodes, while the check nodes in the middle are connected to $6$ variable nodes. The $wb_c=2$ check nodes located at the beginning and at the end of the chain, however, are only connected to either $2$ or $4$ variable nodes. The ensemble design rate for $L>2$ is obtained using (\ref{termrate}) as $$R_L = 1-\frac{L+2}{2L}=\frac{L-2}{2L}.$$ We will denote the SC-LDPC-BC ensemble obtained using this edge spreading and coupling length $L$ as $\mathcal{C}(3,6,L)$. With the exception of the slight structured irregularities at the ends of the graph, the $\mathcal{C}(3,6,L)$ ensembles retain essentially all of the beneficial structural properties of $(3,6)$-regular ensembles with, as will be demonstrated later, dramatically improved iterative BP thresholds.  \hfill $\Box$

In a similar fashion, we obtain $\mathcal{C}(J,K,L)$ SC-LDPC-BC ensembles by terminating the associated $\mathcal{C}(J,K)$ SC-LDPC-CC ensemble with coupling length $L$.

\medskip\noindent{\bf The $\mathcal{C}(J,K,L)$ SC-LDPC-BC Ensemble:} Let $a=\textrm{gcd}(J,K)$ denote the greatest common divisor of $J$ and $K$. Then there exist positive integers $J^\prime$ and $K^\prime$ such that $J=aJ^\prime$ and $K=aK^\prime$ with $\textrm{gcd}(J^\prime,K^\prime)=1$. It follows that the base matrix of a protograph-based SC-LDPC-BC ensemble with coupling width $w=a-1$ can be defined as in (\ref{termbase}), where the submatrices $\mathbf{B}_i$, $i=0,\ldots,w$, are identical $J^\prime\times K^\prime$ matrices with all entries equal to one.  We denote the SC-LDPC-BC ensembles constructed using this edge spreading as $\mathcal{C}(J,K,L)$.\medskip

It follows that $\mathcal{C}_{\mathrm{ARJA}}(L)$ and $\mathcal{C}_{\mathrm{AR4JA}}(e,L)$ SC-LDPC-BC ensembles can be obtained in a similar fashion.

{\bf Example \ref{ex:arja} (cont.)} The irregular ARJA convolutional protograph can be terminated as shown in Fig.~\ref{fig:arja}(b). Note the additional (structured) irregularity introduced when the convolutional protograph is terminated: the variable nodes at each time instant and the check nodes in the center of the chain have the same number and type of connections as the block/convolutional ARJA protograph; however, the check nodes at the ends of the chain have a reduced number of connections. Note also that there are exactly $L$ punctured nodes in the terminated protograph with base matrix $\mathbf{B}_{[0,L-1]}$. As a result of the all-zero row in component matrix $\mathbf{B}_1$ (the disconnected check node in Fig.~\ref{fig:arja}(b)), the terminated protograph associated with $\mathbf{B}_{[0,L-1]}$ has $n_c=(L+w)b_c-1=3L+2$ check nodes and $n_v=Lb_v=5L$ variable nodes. After puncturing, the number of transmitted variable nodes is $n_t=5L-L=4L$ (see Fig.~\ref{fig:arja}(b)) and the design rate of the SC-LDPC-BC ensemble with coupling length $L\geq 2$ is $$R_L=\frac{n_v -n_c}{n_t}=\frac{5L-(3L+2)}{4L}=\frac{L-1}{2L}.$$ In the sequel we will denote the SC-LDPC-BC ensemble obtained using this edge spreading and coupling length $L$ as $\mathcal{C}_{\mathrm{ARJA}}(L)$.

In an identical way, the irregular AR4JA convolutional protograph can be terminated as shown in Fig.~\ref{fig:ar4jacc}. The design rate of the terminated convolutional protograph with extension parameter $e$ and coupling length $L\geq 2$ is given by $$R_L=\frac{(5+2e)L - (3L+2)}{(4+2e)L}=\frac{(1+e)L-1}{(2+e)L}.$$ In the sequel we will denote the SC-LDPC-BC ensemble obtained using this edge spreading and coupling length $L$ as $\mathcal{C}_{\mathrm{AR4JA}}(e,L)$.
\hfill $\Box$

\begin{figure}[t]
\begin{center}
\includegraphics[width=\columnwidth]{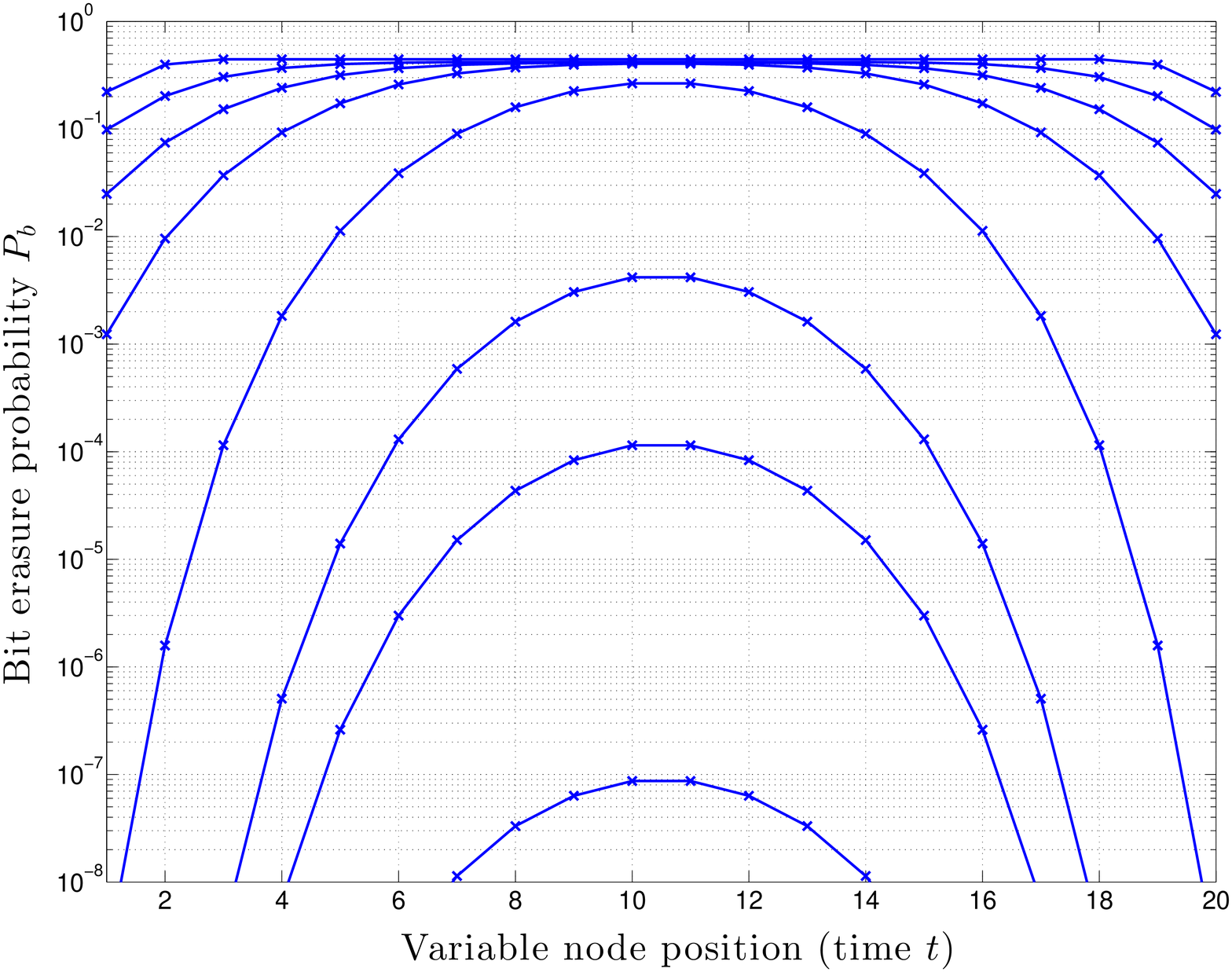} 
\end{center}
\caption{Evolution of the average bit erasure probability $P_b$ of the variable nodes at time $t$ in a code drawn from the $\mathcal{C}(3,6,20)$ SC-LDPC-BC ensemble transmitted over a BEC with erasure probability $\varepsilon=0.48$ for iterations  $i=1,5,20,50, 90, 98, 99, 100$ (from top to bottom).}\label{fig:chainpb}
\end{figure}

In the context of iterative BP decoding, the smaller degree check nodes at the ends of the graph pass more reliable messages to their neighboring variable nodes, and this effect propagates throughout the graph as iterations increase. This effect is demonstrated in Fig.~\ref{fig:chainpb}, where we plot the evolution of the average bit erasure probability $P_b$  on a BEC of the variable nodes at times $t=1,2,\ldots,20$ for a code drawn from the $\mathcal{C}(3,6,20)$ SC-LDPC-BC ensemble with an increasing number of iterations of the BP decoder. We observe that $P_b$ for variable nodes close to the ends of the spatially coupled chain, which are connected to the lower degree check nodes, quickly decreases with iterations, and that this `wave' moves through the chain from either end towards the variable nodes in the centre. In Section~\ref{sec:tradeoff}, we will see that this phenomenon results in excellent iterative decoding thresholds for SC-LDPC-BC ensembles. Note that, after termination, the SC-LDPC-CC ensemble can be viewed as an LDPC-BC ensemble with block length $n=MLb_v$. However, compared to typical LDPC-BC designs that have no restrictions on the location of the ones in the parity-check matrix and hence allow connections across the entire graph, the SC-LDPC-BC ensemble has a highly \emph{localized} graph structure, since the non-zero portion of the parity-check matrix is restricted to a diagonal band of width $\nu$. In addition to the good asymptotic ensemble properties such as excellent BP thresholds and linear minimum distance growth rates that will be demonstrated in Section \ref{sec:tradeoff}, this localized graph structure also gives rise to efficient decoder implementations such as the high-throughput \emph{pipeline decoder} \cite{fz99,pjs+08} and low-latency \emph{sliding window} decoding strategies \cite{lscz10,lpf11,ips+12}.

\subsubsection{Tail-biting LDPC-CC  ensembles}\label{sec:tail}
The convolutional protograph can also be terminated using \emph{tail-biting} \cite{st79,mw86}.

\begin{definition}
\emph{A tail-biting convolutional protograph is obtained from the terminated convolutional protograph with coupling length $L>w$ by combining the check nodes at times $t=L, L+1, \ldots, L+w-1$ with the corresponding check nodes of the same type at times $t=0,1,\ldots,w-1$, respectively. An ensemble of protograph-based \emph{tail-biting spatially coupled LDPC-BCs (TB-SC-LDPC-BCs)} with block length $n=MLb_v$ is then obtained as the collection of all $M$-fold graph covers of the tail-biting convolutional protograph.}
\end{definition}


\noindent The $Lb_c\times Lb_v$ base matrix $\mathbf{B}_{[0,L-1]}^{tb}$ corresponding to the tail-biting convolutional protograph is 

\begin{align}\label{tbbase}
&\mathbf{B}_{[0,L-1]}^{tb} =\nonumber\\&\left[\begin{array}{llllllll}
\mathbf{B}_0 &  & &   &  & \mathbf{B}_{w}   & \cdots &  \mathbf{B}_1 \\
\hspace{1mm}\vdots & \mathbf{B}_0  & &  &    &  & \ddots& \hspace{1mm}\vdots \\
\mathbf{B}_{w-1}\vspace{1mm} & \hspace{1mm}\vdots  &&   &  &  &    & \mathbf{B}_{w}  \\
\mathbf{B}_{w} & \mathbf{B}_{w-1}  & &  &  &  &    & \\
 & \mathbf{B}_{w}  & & \ddots  &  &  &  &  \\
 &   & &  & \mathbf{B}_{0} &  &  &    \\
 &   && \ddots  & \hspace{1mm}\vdots & \mathbf{B}_{0}  &  &  \\
 &   & & \ddots & \mathbf{B}_{w-1}\vspace{1mm}  & \hspace{1mm}\vdots & \ddots &    \\
 &  & && \mathbf{B}_{w} & \mathbf{B}_{w-1} & \cdots &\mathbf{B}_{0}   \\
\end{array}\right],\end{align}
which can be obtained from the terminated base matrix $\mathbf{B}_{[0,L-1]}$ in (\ref{termbase}) by adding the last $wb_c$ rows  to the first $wb_c$ rows. 

Note that the tail-biting protograph has the same design rate \begin{equation}R_L^{tb} =1-L b_c/L b_v=1-b_c/b_v=R,\end{equation} and degree distribution as the convolutional protograph, \emph{i.e.}, there is no structured irregularity introduced to the graph or rate loss after termination. Consequently, tail-biting is a useful way to terminate a convolutional protograph to a block protograph of desired length such that the properties of the convolutional protograph are retained. We denote the TB-SC-LDPC-BC ensemble obtained from the SC-LDPC-BC ensemble $\mathcal{C}(J,K,L)$ with coupling length $L$ as $\mathcal{C}_\mathrm{tb}(J,K,L)$. 

\begin{figure}[t]
\begin{center}
\includegraphics{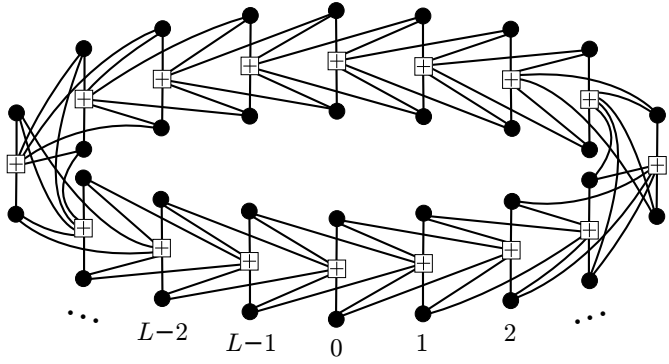} 
\end{center}
\caption{Protograph of a $(3,6)$-regular TB-SC-LDPC-BC ensemble with coupling length $L$ and coupling width $w=2$.}\label{fig:tbring}
\end{figure}

{\bf Example \ref{ex:36proto} (cont.)} Fig.~\ref{fig:tbring} shows the tail-biting $(3,6)$-regular convolutional protograph with coupling length $L$ and coupling width $w=2$.  The corresponding tail-biting base matrix, obtained using (\ref{tbbase}), is
$$\mathbf{B}_{[0,L-1]}^{tb}=\left[\begin{array}{ccccccccccc}\cline{1-2}\cline{8-11}
\multicolumn{1}{|c}{1} & \multicolumn{1}{c|}{1}  &   &    &           &    &     & \multicolumn{1}{|c}{1} & 1 & \multicolumn{1}{|c}{1} & \multicolumn{1}{c|}{1}\\
 \cline{1-4}\cline{8-11}                          
  \multicolumn{1}{|c}{1}    & \multicolumn{1}{c|}{1} & 1 & \multicolumn{1}{c|}{1} &           &    &     &    &    & \multicolumn{1}{|c}{1} & \multicolumn{1}{c|}{1}\\\cline{1-4} \cline{10-11}                                   
\multicolumn{1}{|c}{1} & \multicolumn{1}{c|}{1} & 1 & \multicolumn{1}{c|}{1} & \raisebox{-1.8mm}{$\ddots$} &\\ \cline{1-4}                                 
   &   & \multicolumn{1}{|c}{1} & \multicolumn{1}{c|}{1} & \raisebox{-1.8mm}{$\ddots$} &\\\cline{3-4}\cline{6-7}                                       
   &    &    &    & \raisebox{-1.8mm}{$\ddots$} & \multicolumn{1}{|c}{1} & \multicolumn{1}{c|}{1}\\\cline{6-9}
    &    &    &    &            & \multicolumn{1}{|c}{1} & 1 & \multicolumn{1}{|c}{1} & \multicolumn{1}{c|}{1}\\\cline{6-11} 
    &    &    &    &            & \multicolumn{1}{|c}{1} & 1 & \multicolumn{1}{|c}{1} & 1 & \multicolumn{1}{|c}{1} & \multicolumn{1}{c|}{1}\\\cline{6-11} 
\end{array}\right]_{L\times 2L}.$$

\noindent Note that each variable node has degree $3$ and each check node has degree $6$ in the tail-biting protograph, \emph{i.e.}, the graph is $(3,6)$-regular, the degree distribution is unchanged, and the ensemble design rate is  $R_L^{tb} = 1-L/2L=1/2$. \hfill $\Box$

Protograph-based TB-SC-LDPC-BCs have been used to obtain lower bounds on important parameters of protograph-based SC-LDPC-CC ensembles, such as the free distance \cite{tzc10,mpc13} and minimum trapping set size \cite{mpc13}. 

\subsection{Discussion}\label{sec:scdiscuss}

\subsubsection{Edge Spreading Variations} Given a coupling width $w$, one may construct a convolutional protograph based on a \emph{time-varying edge spreading}, where a different edge spreading is applied at each time instant,  such that
\begin{equation}\label{edgespreadvarying}\sum_{i=0}^w \mathbf{B}_i(t)=\mathbf{B}, \forall t.\end{equation}
(Note that  the degree distribution and computation graphs are not necessarily preserved under this generalization of the Edge Spreading Rule.) Moreover, the construction of a convolutional protograph can be further generalized by coupling a sequence of time-varying protographs, \emph{i.e.}, the base matrix  $\mathbf{B}(t)$ at each time instant depends on $t$.

\subsubsection{Non-Protograph-Based SC-LDPC-CC Construction} Edge spreading can be applied directly to the Tanner graph or parity-check matrix $\mathbf{H}$ of an LDPC-BC to construct an SC-LDPC-CC, without first constructing a block protograph. The two major approaches that have been detailed in the literature can be categorized in this way:

\begin{itemize} 
\item Tanner first developed the connection between \emph{quasi-cyclic (QC)} block codes  
 and time-invariant convolutional codes \cite{tan87}. This approach was extended to construct time-invariant SC-LDPC-CCs in \cite{tss+04}, where the construction can be viewed as a particular infinite graph cover of the Tanner graph of a QC-LDPC-BC \cite{psvc11}; 
\item SC-LDPC-CCs were first introduced in the open literature by Jimenez-Felstr\"{o}m and Zigangirov in 1999 \cite{fz99}. Here, time-varying LDPC-CCs were constructed using a cut-and-paste technique termed \emph{unwrapping}, which is equivalent to applying the Edge Spreading Rule to a sequence of disjoint  LDPC-BC Tanner graphs. As a result of the unwrapping/edge spreading procedure, the computation graph of the underlying LDPC-BC is preserved.\end{itemize} 

\noindent For both construction methods, the SC-LDPC-CCs were shown to have improved BER performance compared to their underlying LDPC-BC counterparts \cite{fz99,tss+04,psvc11}.

\subsubsection{Kudekar's Randomized SC-LDPC-BC Ensemble}\label{sec:kudekar} A  construction of SC-LDPC-BC ensembles, closely related to the $\mathcal{C}(J,K,L)$ ensemble studied in this paper, was presented recently by Kudekar \emph{et al.} \cite{kru11}. Here, $M$ degree $J$ variable nodes and $\frac{J}{K}M$ degree $K$ check nodes are placed at $2L+1$ index positions $[-L,L]$, $L\in \mathbb{N}$. In a similar way to the construction presented here, the graphs are coupled together, where the $J$ connections from the $M$ variable nodes at position $t$ are allowed only to check nodes at positions $[t,t+w-1]$; however, here the connections are \emph{randomized}, such that different edge spreadings are applied to specific fractions of the $M$ variable nodes at each time instant (see \cite{kru11} for a precise definition of the ensemble). This randomized ensemble construction results in  a similar structured irregularity and rate loss due to the boundary (termination) effects as the protograph-based SC-LDPC-BC ensembles presented here. It should be noted, however, that the randomized ensemble and the protograph-based ensemble are mutually exclusive, in the sense that neither code ensemble contains a code from the other ensemble. 


In general, the randomized ensemble does not enjoy as favorable a tradeoff between rate, threshold, and block length as the protograph-based ensemble \cite{kru11}, and it lacks the inherent implementation advantages of a structured ensemble; however, it is a useful ensemble for analytical purposes. In particular, it was shown analytically in \cite{kru11} that the BP threshold for the randomized SC-LDPC-BC ensemble improves all the way to the optimal maximum \emph{a posteriori} (MAP) threshold of the underlying $(J,K)$-regular LDPC-BC ensemble (a fact previously demonstrated numerically for a permutation matrix-based SC-LDPC-CC ensemble in \cite{lscz10}), a phenomenon termed \emph{threshold saturation}. In other words, the randomized ensemble achieves globally optimal decoding performance with low-complexity, locally optimal, iterative BP decoding. In this paper, following the approach applied in \cite{lscz10}, we show numerically that the threshold saturation effect also occurs for carefully designed protograph-based SC-LDPC-BC ensembles. In addition, we show that such ensembles have minimum distance growing linearly with block length, promising excellent performance in both the waterfall \emph{and} error-floor regions of the BER curve.

\subsubsection{Quasi-cyclic Protograph-based Codes} In general, highly structured codes and code ensembles are attractive from an implementation standpoint. In particular, members of a protograph-based LDPC code ensemble that are QC are of great interest to code designers, since they can be encoded with low complexity using simple feedback shift-registers \cite{myk05,lcz+06} and their structure leads to efficiencies in decoder design \cite{wc07b,dyc09}. Moreover, QC-LDPC codes can be shown to perform well compared to randomly constructed LDPC codes for moderate block lengths \cite{klf01,cxdl04,tss+04,khz+10}. The construction of QC-LDPC codes can be seen as a special case of the protograph-based construction in which the $M$-fold graph cover is obtained by restricting the edge permutations to be cyclic, and it can be described by an $Mn_c \times Mn_v$  parity-check matrix formed as an $n_c \times n_v$ array of $M\times M$ circulant matrices. However, unlike typical members of a protograph-based LDPC code ensemble, asymptotic ensemble average results such as iterative decoding thresholds and minimum distance growth rates cannot  be used to describe the behavior of the QC sub-ensemble, since the probability of picking such a code vanishes in the limit of large $M$. For example, if the protograph base matrix consists of only ones and zeros, then the minimum Hamming distance is bounded above by $(n_c+1)!$, where $n_c$ is the number of check nodes in the protograph, regardless of the lifting factor $M$ \cite{md01,fos04}.

\subsubsection{Code Design Flexibility} A nice feature of SC-LDPC-CCs is that, by varying the termination (or coupling) length $L$, we obtain a flexible family of SC-LDPC-BCs with varying rates and frame lengths that display little variation in performance \cite{pjs+08}, \emph{i.e.}, the beneficial properties of spatial coupling are preserved over a range of termination lengths. This is particularly useful in applications or standards that require varying frame lengths, because  one would typically have to design a separate LDPC-BC for each required length. Moreover, if the SC-LDPC-CC is periodic, it is possible to obtain a family of periodically time-varying SC-LDPC-BCs that share the same encoding and decoding architecture for arbitrary $L$.




\section{Minimum Distance and Threshold Trade-Offs for SC-LDPC-BC Ensembles}\label{sec:tradeoff}
In this section, we begin with an asymptotic weight enumerator analysis of protograph-based SC-LDPC-BC ensembles, then proceed by means of a DE analysis to obtain iterative decoding thresholds for both the BEC and AWGNC, demonstrating that the ensembles are both asymptotically good in terms of minimum distance and exhibit the threshold saturation effect with iterative decoding.  

\subsection{Weight Enumerators}\label{sec:weight}
 We begin by summarizing the procedure presented in \cite{ddja09} to obtain the average distance spectrum for a protograph-based ensemble and then apply it to some example SC-LDPC-BC ensembles to test if they are asymptotically good, taking advantage of the fact that the inherent structure of members of a protograph-based LDPC code ensemble facilitates the calculation of average weight enumerators. 
 
Suppose that all $n_v$ variable nodes of the protograph are to be transmitted
over the channel and that each
of the $n_v$ transmitted variable nodes has an associated weight $d_i$, where $0
\leq d_i \leq M$ for all $i$.\footnote{In this context, the `weight' $d_i$ associated with a particular variable node $v_i$ in the protograph refers to the portion of the overall Hamming weight $d$ of a codeword that is distributed over the $M$ variable nodes of type $v_i$ in the $M$-fold graph cover. 
Since we use $M$ copies of the protograph, the weight associated with a
particular variable node in the protograph can be as large as $M$.}
Let $S_d =\{(d_0,d_1,\ldots,d_{n_v-1})\}$ be the set of all possible
weight distributions such that $d_0+d_1+\ldots+d_{n_v-1}=d$. The ensemble average weight enumerator for the protograph is then given by
\begin{equation}
A_d = \sum_{(d_0,d_1,\ldots,d_{n_v-1})\in S_d}A_\mathbf{d},
\end{equation}
where $A_\mathbf{d}$ is the average number of codewords in the
ensemble with a particular weight distribution $$\mathbf{d} = (d_0,d_1,\ldots,d_{n_v-1}).$$
Combinatorial expressions for $A_\mathbf{d}$ have been derived in \cite{ddja09} and \cite{fmt05}. Note that if $n_t<n_v$ variable nodes are to be transmitted over the channel, then the weight enumerator $A_d$ is a double summation over all possible partial weight patterns $S_p$ and $S_d$ of the punctured and transmitted variable node weights, respectively, where the codeword weight $d$ is the sum of the partial weights associated with the transmitted nodes (see \cite{ddja09} for details).

The \emph{asymptotic spectral shape function} of a code
ensemble can be written as $r(\delta) = \lim_{n\rightarrow
\infty}\textrm{sup } r_n(\delta),$ where $r_n(\delta)
=\textrm{ln}(A_d)/{n}$, $\delta = d/n$, $d$ is the Hamming
weight, $n$ is the block length, and $A_d$ is the ensemble average
weight distribution. Suppose that the first positive zero crossing of $r(\delta)$ occurs at $\delta =
\delta_\mathrm{min}$.  If $r(\delta)$ is negative in the range $
0<\delta<\delta_\mathrm{min}$, then $\delta_\mathrm{min}$ is called
the \emph{minimum distance growth rate} of the code ensemble. By
considering the probability
$$\mathbb{P}(d < n\delta_\mathrm{min}) \leq
\sum^{n\delta_\mathrm{min}-1}_{d=1}A_d,$$ it is clear that, as the
block length $n$ becomes sufficiently large, if $\mathbb{P}(d < n\delta_\mathrm{min})
<<1$, then we can say with high probability that a randomly chosen code from the ensemble has a minimum distance that is at least as large as $n\delta_\mathrm{min}$ \cite{ddja09}, i.e.,
 the minimum distance increases linearly with block length $n$. We refer to such an ensemble of codes as \emph{asymptotically good}.
 
 \par\medskip\noindent{\bf Example \ref{ex:36proto} (cont.).} Examining the asymptotic weight enumerators of the $\mathcal{C}(3,6,L)$ ensembles for various coupling lengths $L$, we find that the ensembles are asymptotically good. The calculated minimum distance growth rates are given in Table \ref{tab:36reg}. As the coupling length $L$ tends to infinity, we observe that the minimum distance growth rate $\delta_\mathrm{min}^{(L)}$ decreases. This is consistent with similar results obtained for TB-SC-LDPC-BC ensembles in \cite{mpc13}. We also observe from Table \ref{tab:36reg} that the scaled growth rates $\delta_\mathrm{min}^{(L)}L/(w+1)$ converge to a fixed value as $L$ increases. A similar result was first observed in \cite{stl+07} for an ensemble of $(3,6)$-regular SC-LDPC-CCs constructed from $M \times M$ permutation matrices, where it was shown that the scaled growth rates of the terminated SC-LDPC-BC ensembles converged to a bound on the \emph{free distance} growth rate of the unterminated SC-LDPC-CC ensemble. 
\hfill $\Box$\medskip
\begin{table}[h]
\begin{center}
\caption{Minimum distance growth rates for the $\mathcal{C}(3,6,L)$ SC-LDPC-BC ensembles.}\label{tab:36reg}
\begin{tabular}{|c|c|c|c|}
\hline \hline $L$ &  Design Rate $R_L$& Growth rate $\delta_\mathrm{min}^{(L)}$  & $\delta_\mathrm{min}^{(L)}L/(w+1)$\\[0.5ex]
\hline 
$3$ & $1/6$   & $0.1419$ & $0.142$\\
$4$ & $1/4$   & $0.0814$ & $0.109$\\
$5$ & $3/10$  & $0.0573$ & $0.096$\\
$6$ & $1/3$   & $0.0449$ & $0.090$\\
$7$ & $5/14$  & $0.0374$ & $0.087$\\
$8$ & $3/8$   & $0.0324$ & $0.086$\\
$9$ & $7/18$  & $0.0287$ & $0.086$\\
$10$ & $2/5$  & $0.0258$ & $0.086$\\
$20$ & $9/20$ & $0.0129$ & $0.086$\\
$\infty$ & $1/2$ & $0$ & \\         
\hline\hline 
\end{tabular}
\end{center}

\end{table}

\begin{figure}[t]
\begin{center}
\includegraphics[width=\columnwidth]{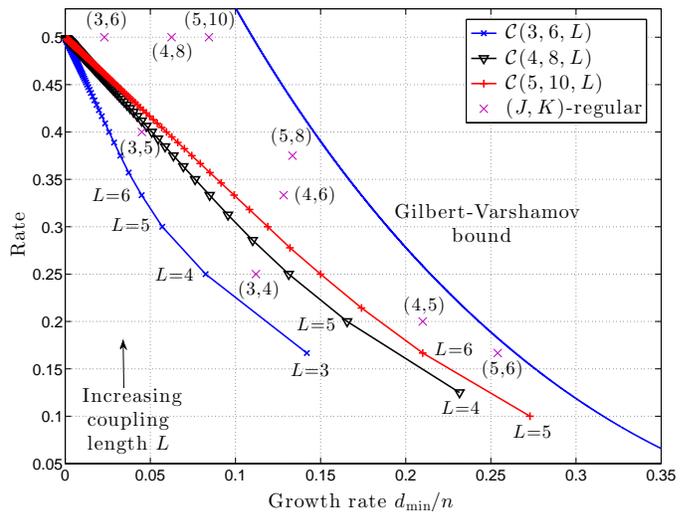}
\end{center}
\caption{Minimum distance growth rates for $\mathcal{C}(J,2J,L)$ SC-LDPC-BC ensembles with design rate $R_L=(L-J+1)/2L$ and some $(J,K)$-regular LDPC-BC ensembles with design rate $R=1 - J/K$. Also shown is the Gilbert-Varshamov bound for random block code minimum distance growth rates.}\label{fig:r12dist}
\end{figure}

In the following example, we consider how the distance growth rates of SC-LDPC-BC ensembles are affected by increasing the density of the graph.
\begin{example}\label{ex:r12}
Consider the $\mathcal{C}(J,2J,L)$ SC-LDPC-BC ensembles. The design rates $R_L$ of these SC-LDPC-BC ensembles  approach the design rates $R=1/2$ of the associated unterminated $\mathcal{C}(J,2J)$ SC-LDPC-CC ensembles as $L\rightarrow \infty$. As we increase the variable node degree $J$, the graph density, and hence the iterative decoding complexity (commonly measured as the average variable and check node degrees), grows. Table \ref{tab:complexity} describes the complexity of the $\mathcal{C}(J,2J,L)$ ensembles. For finite $L$, the average check node degree of the $\mathcal{C}(J,2J,L)$ ensemble is strictly less than $2J$ (the check node degree of the $\mathcal{C}(J,2J)$ SC-LDPC-CC ensemble). The check node degree increases with $L$, tending to $2J$ as $L$ tends to infinity. The variable node degree remains constant at $J$ for all coupling lengths $L$.
\begin{table}[h]
\begin{center}
\caption{Complexity of the $\mathcal{C}(J,2J,L)$ SC-LDPC-BC ensembles.}\label{tab:complexity}
\begin{tabular}{|c|c|c|c|}
 \hline\hline Ensemble    & Design Rate   & Variable  &  Avg. check  \\
               & $R_L$  &  node degree        & node degree       \\
\hline 
$\mathcal{C}(3,6,L)$      & $(L-2)/2L$   & $3$      & $6L/(L+2)$\\
$\mathcal{C}(4,8,L)$        & $(L-3)/2L$   & $4$      & $8L/(L+3)$\\
$\mathcal{C}(5,10,L)$        & $(L-4)/2L$   & $5$      & $10L/(L+4)$\\
$\mathcal{C}(J,2J,L)$ & $(L-J+1)/2L$ & $J$& $2JL/(L+J-1)$\\
\hline\hline
\end{tabular}
\end{center}

\end{table}

Fig.~\ref{fig:r12dist} plots the minimum distance growth rates for $\mathcal{C}(J,2J,L)$ code ensembles with $J = 3,4,$ and $5$, some $(J,K)$-regular LDPC-BC ensembles, and the Gilbert-Varshamov bound \cite{gil52,var57}. As with the $\mathcal{C}(3,6,L)$ ensembles analyzed in Example~\ref{ex:36proto}, we find that the $\mathcal{C}(4,8,L)$ and $\mathcal{C}(5,10,L)$ ensembles are asymptotically good, with large minimum distance growth rates for the lower rate ensembles corresponding to small $L$; then, as the coupling length $L$ is increased, we observe declining minimum distance growth rates as the code rates increase. We again observe that the scaled minimum distance growth rates $\delta_\mathrm{min}^{(L)}L/(w+1)$ converge as $L$ increases, which allows us to estimate the growth rates for $L > 20$ (as explained further in Section~\ref{sec:growth}). As expected, there is a significant increase observed in the growth rates of the $\mathcal{C}(4,8,L)$ ensembles compared to the $\mathcal{C}(3,6,L)$ ensembles of the same rate, and there is a smaller improvement for the $\mathcal{C}(5,10,L)$ ensembles. We would expect this trend to continue as we further increase the variable node degree $J$. \hfill $\Box$
\end{example}

\subsection{Thresholds for the BEC}\label{sec:bec}
In this section, we assume that BP decoding is performed after transmission over a BEC with erasure probability $\varepsilon$. In every decoding iteration, all of the check nodes are updated followed by all of the variable nodes. The messages that are passed between the nodes represent either an erasure or the correct symbol value ($0$ or $1$). For the BEC, a DE analysis of the BP decoder can be performed for an \emph{unstructured} LDPC-BC ensemble with degree distribution pairs $(\lambda(x),\rho(x))$ explicitly by means of the equation 
\begin{equation}\label{de}
p^{(i)} =\varepsilon \lambda\left(1-\rho\left(1-p^{(i-1)}\right)\right),
\end{equation}
where $p^{(i)}$ denotes the probability that a variable to check node message in decoding iteration $i$ corresponds to an erasure, averaged over all codes in the ensemble. Due to this averaging, the message probabilities are equal for all edges in the graph. The DE threshold of an ensemble, defined as the maximum value of the channel parameter $\varepsilon$ for which $p^{(i)}$ converges to zero as $i$ tends to infinity, directly follows from (\ref{de}). Equation (\ref{de}) is also the key to the design of degree distribution pairs $(\lambda(x),\rho(x))$ for capacity achieving sequences of codes with a vanishing gap between the threshold and the Shannon limit (capacity) $\varepsilon_\mathrm{Sh} = 1 - R$ \cite{os02}. Check-concentrated or even check-regular ensembles are known to provide a good trade-off between iterative decoding complexity (measured by the average variable and check node degrees) and gap to capacity.\footnote{Check-concentrated ensembles have a degree distribution such that $\rho(x)$ has a small number of terms. Check-regular ensembles have a degree distribution such that $\rho(x)$ has precisely one term.}

A double exponential decrease of the decoding erasure probability with iterations implies that the probability of erased frames also converges to zero \cite{ltzc05}. The lower bounds in \cite{gws11} on the decoding complexity of general message passing decoders, obtained using sphere-packing arguments, also predict a double exponential reduction of the error (erasure) probability with the number of iterations. A Taylor expansion of (\ref{de}) reveals that the erasure probability $p^{(i)}$ converges to zero at least doubly exponentially with $i$ if all nodes have a variable node degree of at least three. An analysis by means of the messages' Bhattacharyya parameter shows that this is also true for general MBS channels \cite{ltzc05}. Unstructured irregular LDPC-BC ensembles with thresholds close to capacity exhibit a non-vanishing fraction of degree two variable nodes, and thus their frame error (erasure) probabilities are bounded away from zero; however, {structured protograph-based LDPC code ensembles containing degree two variable nodes can achieve the desired double exponential decay} \cite{pst13}{. Protograph-based $\mathcal{C}(J,K,L)$ ensembles achieve this double exponential decay in error probability for $J\geq 3$.}

Since every member of a protograph-based ensemble preserves the structure of the base protograph, DE analysis for the resulting codes can be performed within the protograph. We now describe the application of DE to \emph{structured} protograph-based ensembles. It is useful to label the edges in $E$ from both a variable node and a check node perspective. Then $e^{v}_{y,l}$ indicates the $l$th edge emanating from variable node $v_y$. Similarly, $e^{c}_{x,m}$ denotes the $m$th edge emanating from check node $c_x$. Note that $l \in \{1,\ldots, \partial(v_y)\}$ and $m \in \{1,\ldots, \partial(c_x)\}$, where $\partial(v_y)$ and $\partial(c_x)$ denote the degree of variable node $v_y$ and check node $c_x$, respectively. It follows that if $e^{v}_{y,l}$ and $e^{c}_{x,m}$ define the same edge, $v_y$ is connected to $c_x$.

For a BEC, let $q^{(i)}(e^{c}_{x,m})$ denote the probability that the check to variable node message sent along edge $e^{c}_{x,m}$ in decoding iteration $i$ is an erasure. (Note that this will be the case if at least one of the incoming messages from other neighboring variable nodes is erased.) Explicitly, 
\begin{equation}\label{deq}
 q^{(i)}\left(e^{c}_{x,m}\right)=1 - \prod_{m^\prime \neq m}\left(1-p^{(i-1)}\left(e^{c}_{x,m^\prime}\right)\right),
\end{equation}
where $p^{(i-1)}(e^{c}_{x,m^\prime})$ denotes the probability that the incoming message in the previous update of check node $x$ is an erasure and $m$, $m^\prime \in \left\{1,\ldots,\partial(c_x)\right\}$. In contrast, the variable to check node message sent along edge $e^{v}_{y,l}$ is an erasure if the incoming message from the channel and the messages from all the other neighboring check nodes are erasures. This happens with probability $p^{(i)}(e^{v}_{y,l})$, where
\begin{equation}\label{dep}
 p^{(i)}\left(e^{v}_{y,l}\right)=\varepsilon\prod_{l^\prime \neq l}q^{(i)}\left(e^{v}_{y,l^\prime}\right)
\end{equation}
and $l$, $l^\prime \in \left\{1,\ldots,\partial(v_y)\right\}$. The \emph{BP decoding threshold} $\varepsilon^*$ of a protograph-based ensemble is defined as the maximum value of the channel parameter $\varepsilon$ for which $p^{(i)}(e^{v}_{y,l})$ converges to zero as $i$ tends to infinity for all edges $e^{v}_{y,l}$ emanating from variable node $v_y$ and for all variable nodes ${v}_{y}$ in the protograph.

 \par
\begin{figure}[t]
\begin{center}
\includegraphics[width=\columnwidth]{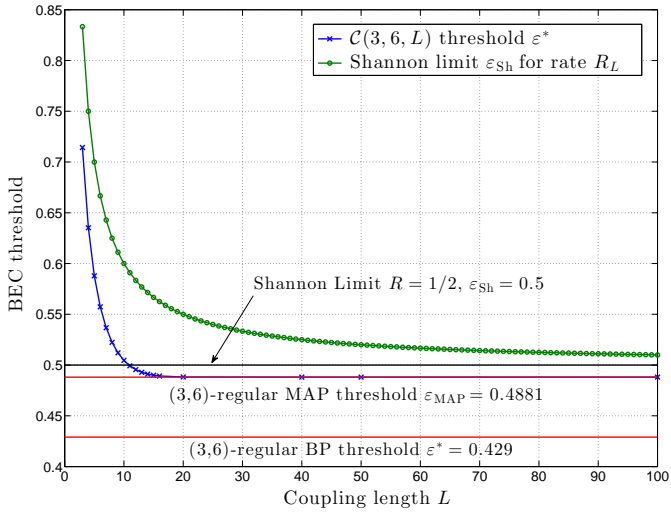}
\end{center}
\caption{BEC iterative BP decoding thresholds for $\mathcal{C}(3,6,L)$ SC-LDPC-BC ensembles with design rate $R_L=(L+2)/2L$ and the corresponding Shannon limit $\varepsilon_{\mathrm{Sh}} = 1-R_L$ for rate $R_L$. Also shown are the BP and MAP decoding thresholds for the underlying $(3,6)$-regular LDPC-BC ensemble, $\varepsilon^*=0.429$ and $\varepsilon_{\mathrm{MAP}}=0.4881$, respectively, and the Shannon limit for $R=1/2$ codes, $\varepsilon_{\mathrm{Sh}}=0.5$.}\label{fig:r12thres}
\end{figure}

\medskip\noindent{\bf Example \ref{ex:36proto} (cont.).} 
 Fig.~\ref{fig:r12thres} shows the  calculated BP decoding thresholds $\varepsilon^*$ obtained for the $\mathcal{C}(3,6,L)$ SC-LDPC-BC ensembles by recursive application of (\ref{deq}) and (\ref{dep}) for different channel parameters $\varepsilon$. Also shown is the corresponding Shannon limit $\varepsilon_{\mathrm{Sh}}=1-R_L$. For small values of $L$, where the design rate is lower, we observe large thresholds (\emph{e.g.}, $\varepsilon^* =0.635$ for $L=3$, where $R_3 =1/4$ and $\varepsilon_{\mathrm{Sh}} =0.750$, resulting in a gap to capacity of $0.115$). As we increase $L$, the rate increases and the thresholds decrease; however, the gap to capacity also decreases (\emph{e.g.}, $\varepsilon^* =0.505$ for $L=10$, where $R_{10} =2/5$ and $\varepsilon_{\mathrm{Sh}} =0.6$, resulting in a gap to capacity of $0.095$). When $L$ becomes sufficiently large (in this example around $L=20$), the threshold converges, or \emph{saturates}, to a constant value $\varepsilon^* = 0.488$. As $L$ is further increased and the rate approaches $R_\infty = 1/2$, the threshold remains constant at $\varepsilon^* = 0.488$, \emph{i.e.}, it displays the remarkable property that it does not continue to decay as the design rate of the ensembles increases and approaches $R_\infty = 1/2$. The Shannon limit is equal to $\varepsilon_{\mathrm{Sh}}  = 0.5$ for rate $R_\infty = 1/2$, and thus the gap to capacity  decreases with increasing $L$ to the constant value $0.012$. Since the gap to capacity improves with increasing $L$, while the distance growth rate worsens with $L$ (see Table \ref{tab:36reg}), this indicates the existence of a trade-off between distance growth rate and threshold.
 
Also shown in Fig.~\ref{fig:r12thres} are the sub-optimal, low complexity BP threshold $\varepsilon^*=0.429$ and the optimal, high complexity MAP decoding threshold $\varepsilon_{\mathrm{MAP}}=0.4881$ for the underlying $(3,6)$-regular LDPC-BC ensemble. (Note that even with optimal decoding, there is still a small gap to capacity for a $(3,6)$-regular LDPC-BC ensemble.) We observe that the BP thresholds of the $\mathcal{C}(3,6,L)$ SC-LDPC-BC ensembles are significantly larger than the BP threshold of a $(3,6)$-regular LDPC-BC ensemble for all $L$. Moreover, the BP thresholds of the $\mathcal{C}(3,6,L)$ ensembles converge to a value numerically indistinguishable from the MAP decoding threshold of a $(3,6)$-regular LDPC-BC ensemble, \emph{i.e.}, threshold saturation is observed. Recall that as $L\rightarrow \infty$, $R_\infty =1/2$ and the $\mathcal{C}(3,6,L)$ ensemble degree distribution approaches $(3,6)$-regular; consequently, the $\mathcal{C}(3,6,L)$ ensemble displays the remarkable property of achieving optimal decoding performance with low complexity BP decoding!  As we will observe in the remainder of this section, this phenomenon occurs for all of the protograph-based $\mathcal{C}(J,K,L)$ ensembles. Indeed, it has recently been proven analytically that the BP thresholds of the randomized $\mathcal{C}(J,K,L)$ ensemble described in Section~\ref{sec:kudekar} saturate precisely to the MAP decoding thresholds of their underlying $(J,K)$-regular LDPC-BC ensembles, both for the BEC \cite{kru11} and for general MBS channels \cite{kru13}. Finally, we note that, in conjunction with the excellent thresholds, all variable nodes in the $\mathcal{C}(3,6,L)$   ensembles have degree greater than two and thus, asymptotically, the error probability converges at least doubly exponentially with decoding iterations. \hfill $\Box$\medskip

The interesting phenomenon that the calculated thresholds do not decay as $L$ increases beyond a certain value was first observed empirically in \cite{slcz04} for $\mathcal{C}(J,2J,L)$ SC-LDPC-BC ensembles constructed from $M \times M$ permutation matrices, and it was shown to be true for arbitrarily large $L$ in \cite{lscz10}.  To prove this result, a sliding window updating schedule can be considered, where the decoder updates the nodes only within a window of size $W \leq L$, starting at time $t = 0$. Once the variable-to-check node message probabilities $p(e_{y,l}^v )$, $l = 1,2,\ldots,\partial(v_y)$, are below some value $\varepsilon_0$ for all nodes $v_y$, $y=0,1,,\ldots,b_v-1$, at time $t$, the window is shifted one time unit further.  Suppose that the message probabilities at times $t<0$ are initialized by some value $\varepsilon_0>0$. If, under these conditions, the value $\varepsilon_0$ is reached at time $t = 0$ after some number of iterations, so that the window can be shifted one step further, then, for the actual initial probabilities $p(e^v_{y,l} ) = 0$ of nodes at times $t < 0$, the value $\varepsilon_0$ can also be reached at all times $t$, $t = 0,\ldots,L-1$.

Intuitively, one can explain the result as follows: during the iterations, due to the lower check node degrees at the start of the graph, the messages along edges at time $t = 0$ will be the most reliable ones. Their erasure probabilities thus have the potential to converge to zero even for channel parameters $\varepsilon$ beyond the threshold of the underlying LDPC-BC ensemble. But when the symbols at $t = 0$ are perfectly known, the connected edges can be removed from the protograph with base matrix $\mathbf{B}_{[0,L-1]}$, which results in a shortened protograph with base matrix $\mathbf{B}_{[1,L-1]}$. It follows now by induction that the messages eventually converge to zero at all times $t = 0,\ldots,L-1$ for an arbitrary coupling length $L$.

 \par
\begin{figure}[t]
\begin{center}
\includegraphics[width=\columnwidth]{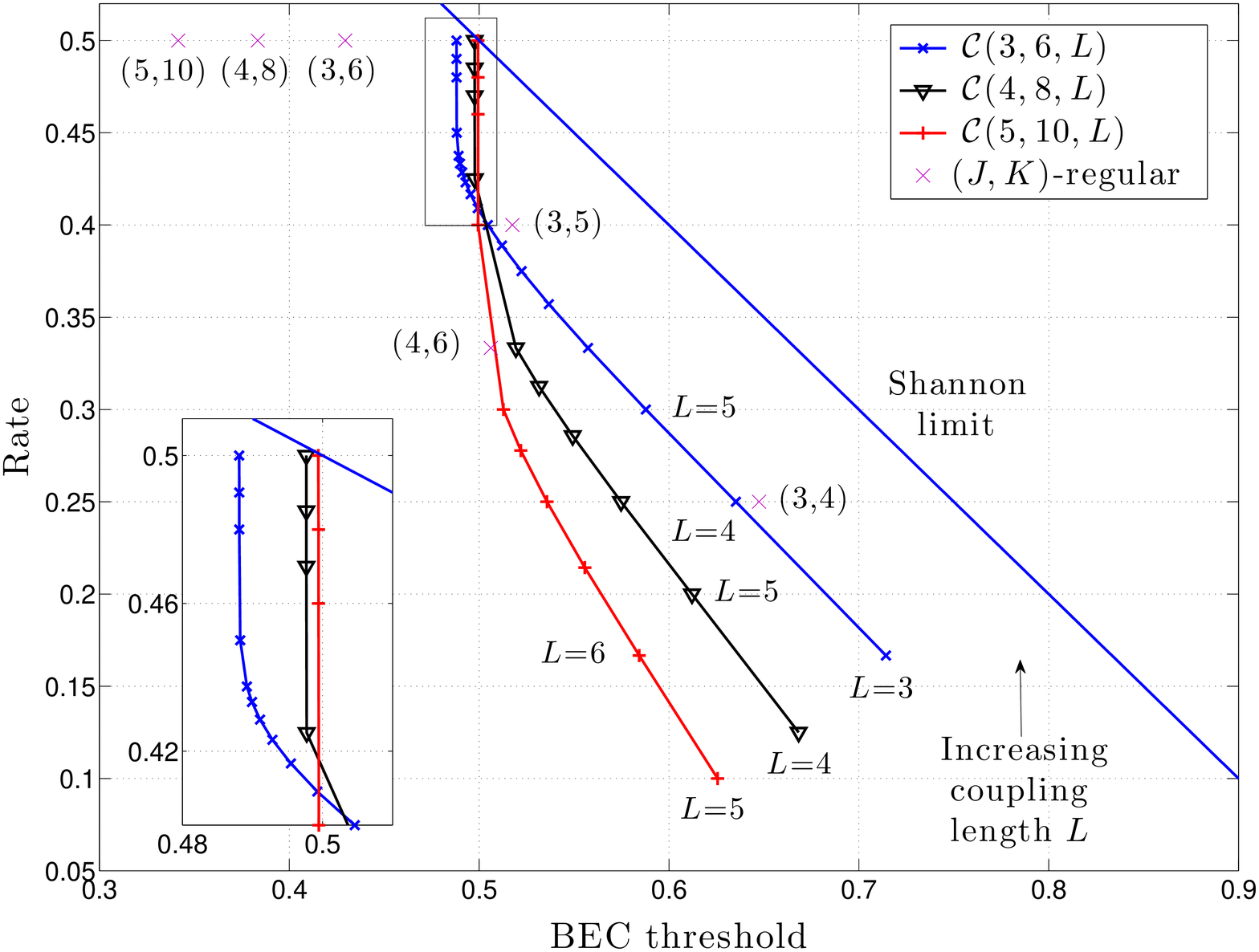}
\end{center}
\caption{BEC iterative BP decoding thresholds for $\mathcal{C}(J,2J,L)$ SC-LDPC-BC ensembles with design rate $R_L=(L-J+1)/2L$ and some $(J,K)$-regular LDPC-BC ensembles with design rate $R=1 - J/K$.}\label{fig:r12thresh}
\end{figure}

\medskip\noindent{\bf Example \ref{ex:r12} (cont.).} 
Fig.~\ref{fig:r12thresh} shows the BEC iterative decoding thresholds for several $\mathcal{C}(J,2J,L)$ SC-LDPC-BC ensembles.  In each case, we observe that the gap to capacity decreases as the coupling length $L$ increases. Also, for a fixed rate and small values of $L$, we see that the thresholds worsen as we increase $J$, which is consistent with the behavior observed for fixed rate $(J,2J)$-regular LDPC-BC ensembles, \emph{e.g.}, the $(3,6)$-, $(4,8)$-, and $(5,10)$-regular ensembles shown in the figure. Recall from Fig.~\ref{fig:r12dist} that,  for a fixed rate and small values of $L$, the minimum distance growth rates  of the $\mathcal{C}(J,2J,L)$ ensembles improve as we increase $J$, which is also consistent with the behavior observed for fixed rate $(J,2J)$-regular LDPC-BC ensembles. Thus, in the small $L$ regime, SC-LDPC-BC ensembles behave like LDPC-BC ensembles, \emph{i.e.}, thresholds worsen and distance growth rates improve by increasing $J$  (and hence iterative decoding complexity).

However, as $L$ increases,  the thresholds of the $\mathcal{C}(J,2J,L)$ ensembles each saturate to a value numerically indistinguishable from the MAP decoding threshold (and significantly larger than the BP threshold) of the underlying $(J,2J)$-regular LDPC-BC ensemble. These values \emph{improve}, rather than worsen, as we increase $J$ ($\varepsilon^*=0.4881$, $0.4977$, and $0.4994$ for the $\mathcal{C}(3,6,L)$, $\mathcal{C}(4,8,L)$, and $\mathcal{C}(5,10,L)$ ensembles, respectively). This indicates that, unlike the underlying $(J,2J)$-regular LDPC-BC ensembles, for large $L$, both the distance growth rates \emph{and} the BP thresholds improve with increasing complexity, and we would expect this trend to continue as we further increase the variable node degree $J$.  Moreover, as we let $J\rightarrow \infty$, the MAP threshold (for an arbitrary MBS channel) of the underlying $(J,2J)$-regular LDPC-BC ensemble improves all the way to the Shannon limit \cite{mb01}. This allows the construction of \emph{capacity achieving} $\mathcal{C}(J,2J,L)$ SC-LDPC-BC ensembles with BP decoding as the graph density grows unbounded.\footnote{Here we observe numerically that the BP thresholds of the $\mathcal{C}(J,2J,L)$ ensembles saturate to the MAP threshold of the underlying $(J,K)$-regular LDPC-BC ensemble. Hence these ensembles are not capacity achieving in the strict sense.  However, the randomized $\mathcal{C}(J,2J,L)$ ensembles in \cite{kru11,kru13} are provably capacity achieving in this regard.} \hfill $\Box$\medskip

\begin{figure*}[t]
\begin{center}
\includegraphics[width=7.1in]{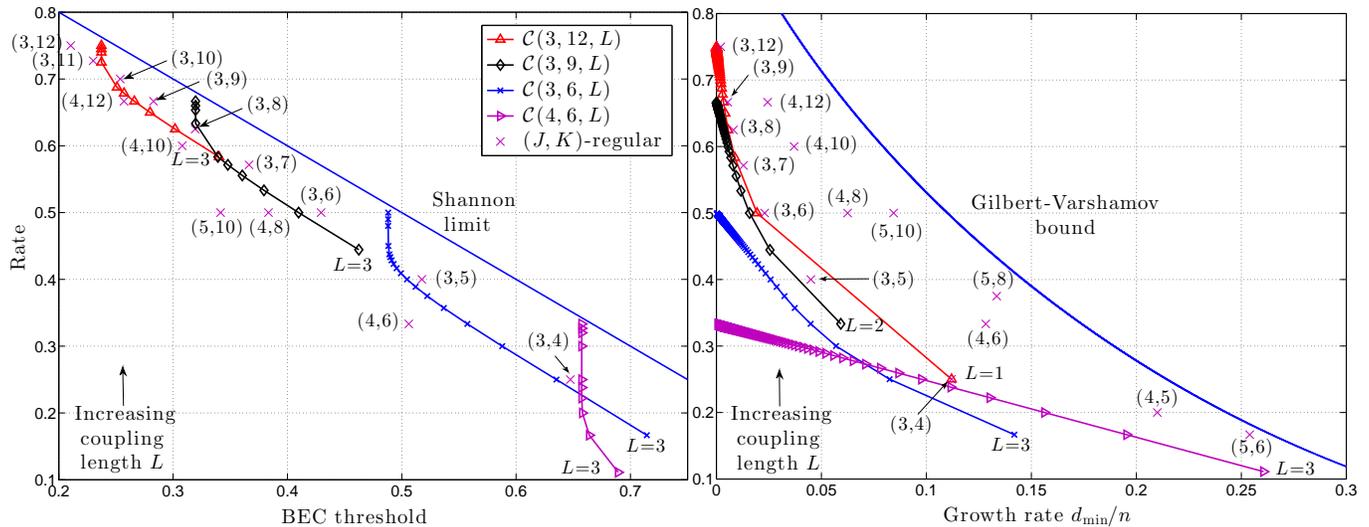}
\end{center}
\caption{BEC iterative BP decoding thresholds and minimum distance growth rates of four $\mathcal{C}(J,K,L)$ SC-LDPC-BC ensembles and several uncoupled $(J,K)$-regular LDPC-BC ensembles.}\label{fig:JK}
\end{figure*}

The results described so far are indicative of the general behavior of $\mathcal{C}(J,K,L)$ SC-LDPC-BC ensembles. In practice, the coupling length $L$ adds an extra degree of freedom. We obtain a family of asymptotically good ensembles with varying iterative decoding thresholds and minimum distance growth rates covering a wide variety of design rates. Moreover, the desired range of achievable SC-LDPC-BC design rates can be extended by coupling higher or lower rate LDPC-BC protographs together using the Edge Spreading Rule.
\begin{example}\label{ex:rates}
In this example, we consider the $\mathcal{C}(3,12,L)$, $\mathcal{C}(3,9,L)$, $\mathcal{C}(3,6,L)$, and $\mathcal{C}(4,6,L)$ SC-LDPC-BC ensembles. Each ensemble has design rate approaching $R_\infty=1-J/K$ (the rate of the underlying $(J,K)$-regular LDPC-BC ensemble), with the usual structured irregularity occurring as a result of the termination. Fig.~\ref{fig:JK} displays the BEC thresholds and distance growth rates of these $\mathcal{C}(J,K,L)$ ensembles and, for comparison, several uncoupled $(J,K)$-regular LDPC-BC ensembles, along with the Shannon limit and the Gilbert-Varshamov bound, respectively. For each family, when $L$ is small and the design rate is low, the iterative decoding thresholds are further from capacity and the minimum distance growth rates are larger compared to the ensembles with larger $L$. Then, as $L$ increases, the gap to capacity decreases and the BP threshold saturates to a value close to the Shannon limit (\emph{i.e.}, the MAP decoding threshold of the underlying $(J,K)$-regular LDPC-BC ensemble)  and significantly better than the BP threshold of the underlying $(J,K)$-regular LDPC-BC ensemble. It follows that, as in the previous  examples, we observe a minimum distance vs. threshold trade-off for each of these  $\mathcal{C}(J,K,L)$ ensembles, since both the minimum distance growth rates and the gap to capacity decrease with increasing $L$. Finally, we note that the design rates $R_L$ of the $\mathcal{C}(J,K,L)$ ensembles included in Fig.~\ref{fig:JK}, given by (\ref{termrate}), cover a wide range of values. \hfill $\Box$
\end{example}

The choice of edge spreading affects the properties of the SC-LDPC-BC ensembles, in particular for small to moderate values of $L$. In the next example, we will see that it is possible to improve both the minimum distance growth rates and thresholds simultaneously by carefully selecting the edge spreading.
\begin{example}\label{ex:edgespreadings}
\begin{figure}[t]
\begin{center}
\includegraphics{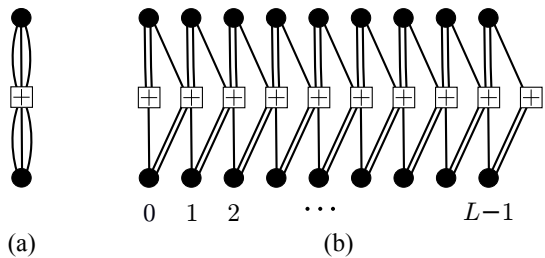} 
\end{center}
\caption{Protographs of (a) a $(3,6)$-regular LDPC-BC ensemble, and (b) a resulting SC-LDPC-BC ensemble with coupling length $L$ obtained by applying the Edge Spreading Rule with coupling width $w=1$.}\label{fig:multi36}
\end{figure}
In this example, we consider a different edge spreading than that chosen for the $\mathcal{C}(3,6,L)$ ensembles. Fig.~\ref{fig:multi36}(a) shows the $(3,6)$-regular protograph used to construct the $\mathcal{C}(3,6,L)$ ensembles. This protograph is copied $L$ times and an edge spreading with coupling width $w=1$ is applied as shown in Fig.~\ref{fig:multi36}(b). We will denote the SC-LDPC-BC ensembles obtained with this edge spreading as $\mathcal{C}_A(3,6,L)$. Comparing the $\mathcal{C}_A(3,6,L)$ ensembles to the $\mathcal{C}(3,6,L)$ ensembles, we notice two major differences. Structurally, the $\mathcal{C}_A(3,6,L)$ ensembles are more regular. The first and last check nodes have degree $3$ and all the other check nodes have degree $6$. Secondly, since $w=1$, the ensemble design rate  $R_L=(L-1)/2L$ obtained using (\ref{termrate}) is larger for a given $L$, \emph{i.e.}, there is less rate loss. Asymptotically in $L$, both ensembles approach $R_\infty = 1/2$ and a $(3,6)$-regular degree distribution.

The BEC thresholds and distance growth rates calculated for the $\mathcal{C}(3,6,L)$ and $\mathcal{C}_A(3,6,L)$ ensembles are displayed in Table~\ref{tab:comp}. We observe that, even though the $\mathcal{C}_A(3,6,L)$ ensembles have a very small structured irregularity (only one reduced degree check node at either end of the chain), like the $\mathcal{C}(3,6,L)$ ensembles their BP thresholds still saturate to the optimal MAP decoding threshold of the underlying $(3,6)$-regular LDPC-BC ensemble. Moreover, the $\mathcal{C}_A(3,6,L)$ ensembles are also asymptotically good and display \emph{both} larger growth rates and better thresholds than the $\mathcal{C}(3,6,L)$ ensembles. 
\begin{table}[t]
\begin{center}
\caption{BEC thresholds and distance growth rates for SC-LDPC-BC ensembles obtained by two different edge spreadings of a $(3,6)$-regular LDPC-BC protograph.}\label{tab:comp}
\begin{tabular}{|c|c|c|c|c|c|c|c|c|c|}\hline\hline
 Design Rate&  \multicolumn{2}{c|}{$\mathcal{C}(3,6,L)$}& \multicolumn{2}{c|}{$\mathcal{C}_A(3,6,L)$}\\\hline
 $R_L$ &  $\varepsilon^*$ & $\delta_\mathrm{min}^{(L)}$ & $\varepsilon^*$ & $\delta_\mathrm{min}^{(L)}$\\\hline
 $1/4$& $0.6353$  & $0.0814$ &   $0.6448$  &  $0.0950$   \\
$1/3$ & $0.5574$  & $0.0449$&  $0.5671$  &  $0.0524$   \\
$3/8$ & $0.5223$  & $0.0324$ &   $0.5301$  &  $0.0375$  \\
 $2/5$ & $0.5046$  & $0.0258$ &  $0.5103$  &  $0.0298$  \\
 $5/12$ & $0.4955$  & $0.0215$&   $0.4993$  &  $0.0248$   \\
 $3/7$ & $0.4911$  & $0.0184$&   $0.4933$  &  $0.0213$   \\
$7/16$  & $0.4892$  & $0.0161$&   $0.4903$  &  $0.0186$  \\
 $19/40$ &$0.4881$  & $0.0065$& $0.4881$ & $0.0074$ \\
$1/2$ &$0.4881$ & $0$&$0.4881$ & $0$ \\
\hline\hline
\end{tabular}
\end{center}

\end{table}

The larger distance growth rates obtained for the $\mathcal{C}_A(3,6,L)$ ensembles can be attributed to having no degree $2$ check nodes and a larger proportion of non-zero elements in  $\mathbf{B}_{[0,L-1]}$, \emph{i.e.}, a denser base matrix. We note that by retaining some repeated edges in the  $\mathcal{C}_A(3,6,L)$ ensemble protograph, the memory requirements for decoder implementation are reduced, \emph{i.e.}, the  SC-LDPC-CC constraint length is $\nu=M(w+1)b_v=4M$, compared to $\nu=M(w+1)b_v=6M$ for the $\mathcal{C}(3,6,L)$ ensembles. Moreover, constructing SC-LDPC-CC ensembles from protographs with repeated edges in order to reduce memory requirements has been shown to improve the performance of a windowed decoder \cite{ips+12}. \end{example}\hfill $\Box$


There are many ways of spreading the edges among the component submatrices $\mathbf{B}_i$ of a base matrix $\mathbf{B}$, and different constructions can result in varying thresholds and ensemble growth rates. (For some other examples of different $(3,6)$-regular edge spreadings  see \cite{lmfc10,mslc10}.) Choices containing all-zero rows and/or columns in the submatrices should be avoided, since they can lead to disconnected subgraphs. Simple row and column permutations (applied to all component submatrices) do not affect the graph structure, and so, in turn, they do not affect the threshold and distance growth rate of the ensemble. A good threshold is expected when the check nodes at the ends of the graph have low degree (but at least degree $2$). This gives an initial convergence boost to the iterative decoder, and the spatially coupled structure allows this reliable information generated at the ends of the graph to propagate through the chain to the centre.

We conclude this section by investigating the minimum distance and threshold trade-off for the irregular $\mathcal{C}_\mathrm{ARJA}(L)$ and $\mathcal{C}_\mathrm{AR4JA}(L)$ SC-LDPC-BC ensembles.
 
 \par\medskip\noindent{\bf Example \ref{ex:arja} (cont.).} The minimum distance growth rates and BEC iterative decoding thresholds for the $\mathcal{C}_\mathrm{ARJA}(L)$ ensembles are given in Table \ref{tab:arja}. (For reference, the underlying ARJA LDPC-BC ensemble has minimum distance growth rate $\delta_\mathrm{min}=0.0145$ and BEC threshold $\varepsilon^*=0.4387$.) Similar to the $\mathcal{C}(J,K,L)$ ensembles, we observe that the ensembles are each asymptotically good; but as the coupling length $L\rightarrow \infty$, the minimum distance growth rate $\delta_\mathrm{min}^{(L)}\rightarrow 0$. We also observe from Table~\ref{tab:arja} that the scaled growth rates $n_t\delta_\mathrm{min}^{(L)}$ converge  as $L$ increases. We will see in Section~\ref{sec:growth} that the scaled growth rates converge to a bound on the \emph{free distance} growth rate of the unterminated $\mathcal{C}_\mathrm{ARJA}$ ensemble. This allows us to estimate the minimum distance growth rate $\delta_\mathrm{min}^{(L)}$ of the $\mathcal{C}_\mathrm{ARJA}(L)$ ensembles for large $L$ by dividing this value by the number of transmitted nodes in the protograph $n_t=4L$. 

\begin{table}[h]
\begin{center}
\caption{Distance growth rates and BEC  thresholds for the ARJA SC-LDPC-BC ensembles $\mathcal{C}_\mathrm{ARJA}(L)$.}\label{tab:arja}
\begin{tabular}{|c|c|c|c|c|c|c|}
 \hline\hline $L$ & Rate & Growth  & Scaled & BEC & Capacity & Gap to\\
  & $R_L$ & rate $\delta_\mathrm{min}^{(L)}$ & $n_t\delta_\mathrm{min}^{(L)}$ & threshold & $\varepsilon_{sh}$&Capacity \\
\hline $2$ & $1/4$ & $0.0946$ & $0.757$ & $0.6608$ & $0.7500$& $0.0892$\\
$3$ & $1/3$   & $0.0461$ & $0.553$ & $0.5864$ & $0.6667$& $0.0803$\\
$4$ & $3/8$   & $0.0306$ & $0.490$ & $0.5496$ & $0.6250$& $0.0750$\\
$5$ & $2/5$   & $0.0234$ & $0.469$ & $0.5284$ & $0.6000$& $0.0716$\\
$6$ & $5/12$  & $0.0192$ & $0.462$ & $0.5159$ & $0.5833$& $0.0674$\\
$7$ & $3/7$   & $0.0164$ & $0.461$ & $0.5083$ & $0.5714$& $0.0631$\\
$8$ & $7/16$  & $0.0144$ & $0.461$ & $0.5039$ & $0.5625$& $0.0586$\\
$9$ & $4/9$   & $0.0128$ & $0.461$ & $0.5016$ & $0.5556$& $0.0540$\\
$10$ & $9/20$ & $0.0115$ & $0.461$ & $0.5004$ & $0.5500$& $0.0496$\\
$\infty$ & $1/2$ & $0$ &            & $0.4996$ & $0.5000$& $0.0004$\\
\hline\hline
\end{tabular}
\end{center}

\end{table}

In addition, we see the same type of threshold behavior exhibited by the $\mathcal{C}(J,K,L)$ ensembles. The BEC iterative decoding threshold saturates to 
$\varepsilon^* = 0.4996$ as $L$ becomes sufficiently large and does not further decay as $L\rightarrow\infty$. This is very close to the Shannon limit $\varepsilon_{sh}=0.5$ for rate $R_\infty=1/2$ and is significantly larger than the BP threshold of the ARJA LDPC-BC ensemble. As the coupling length $L$ increases, we also observe that the gap to capacity decreases, resulting in the usual trade-off between distance growth rate and threshold.

\begin{figure*}[t]
\begin{center}
\includegraphics[width=6.7in]{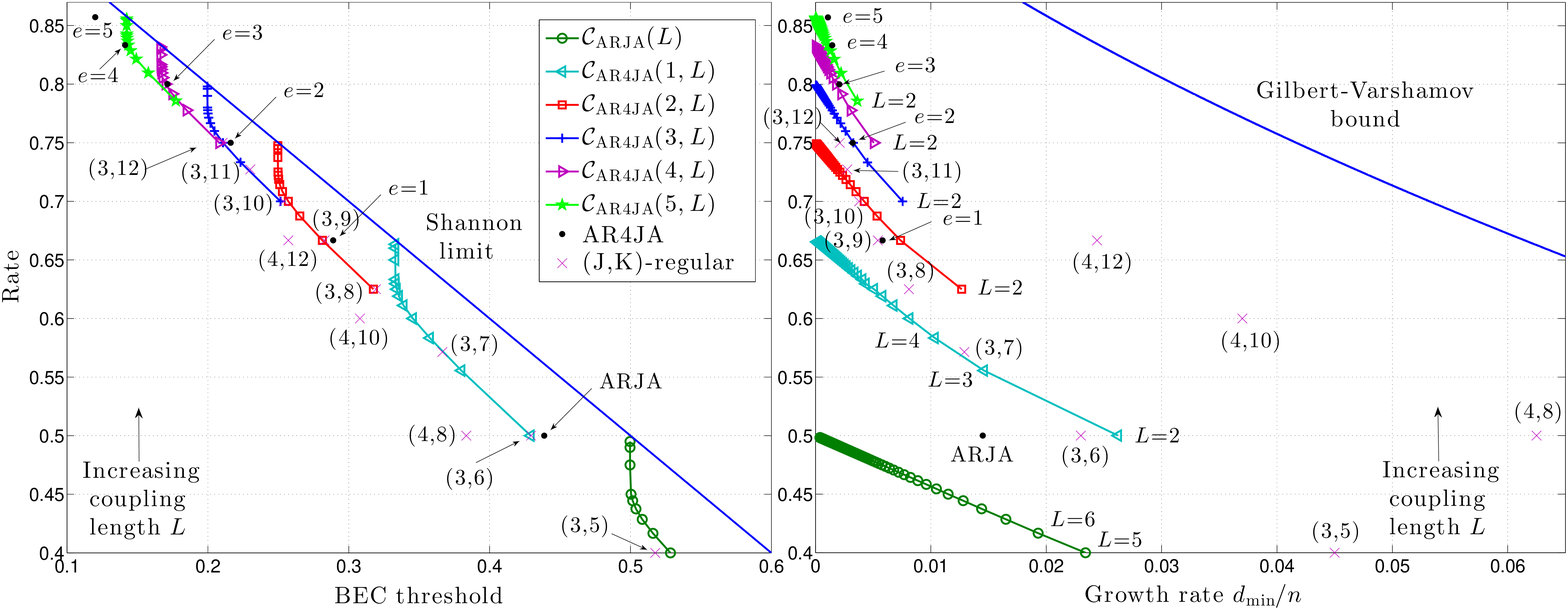}
\end{center}
\addtocounter{figure}{2}
\caption{BEC iterative BP decoding thresholds and minimum distance growth rates for the $\mathcal{C}_\mathrm{AR4JA}(e,L)$ SC-LDPC-BC ensembles, the underlying AR4JA LDPC-BC ensembles, and several $(J,K)$-regular LDPC-BC ensembles.}\label{fig:arjacompzoom}
\end{figure*}


Fig.~\ref{fig:arjacompzoom} shows the results obtained for the $\mathcal{C}_\mathrm{AR4JA}(e,L)$ SC-LDPC-BC ensembles, the underlying AR4JA LDPC-BC ensembles, and several $(J,K)$-regular LDPC-BC ensembles. For the $\mathcal{C}_\mathrm{AR4JA}(e,L)$ ensembles with $e=1,\ldots,5$, we observe that, as in the $e=0$ case, increasing the coupling length $L$ results in asymptotically good code ensembles with capacity approaching iterative decoding thresholds and declining minimum distance growth rates. For each family, the iterative decoding threshold converges to a value close to the Shannon limit for $R_{\infty}$ (and significantly larger than the BP threshold of the underlying AR4JA LDPC-BC ensemble) as $L$ gets large. The design rates $R_L$ of the $\mathcal{C}_\mathrm{AR4JA}(e,L)$ ensembles overlap for increasing extension parameter $e$, allowing a large selection of asymptotically good codes to be obtained in the rate range $1/4 \leq R \leq 6/7$, and the achievable code rate can be increased further by considering larger extension parameters $e$. 

We also observe that the minimum distance growth rates of the $\mathcal{C}_\mathrm{AR4JA}(e,L)$ ensembles for small coupling lengths $L$ typically exceed those of $(3,K)$-regular codes for $K \geq 6$. Further, for the same extension parameter $e$ and large $L$, the $\mathcal{C}_\mathrm{AR4JA}(e,L)$ ensembles have significantly better thresholds and less complexity than the underlying AR4JA LDPC-BC ensembles,\footnote{Complexity, as earlier, is measured by average variable and check node degrees. When comparing the $\mathcal{C}_\mathrm{AR4JA}(e,L)$ ensembles to the underlying AR4JA LDPC-BC ensembles with equal extension parameters, the average variable node degree is the same for all $L$, but the average check node degree is less for the $\mathcal{C}_\mathrm{AR4JA}(e,L)$ ensembles because of the termination.} but smaller distance growth rates and slightly lower code rates. Further, by increasing the extension parameter $e$, and for small $L$, the minimum distance growth rates of the $\mathcal{C}_\mathrm{AR4JA}(e,L)$ ensembles are larger than those of the AR4JA ensemble with only a slightly worse threshold and some increase in complexity.

\begin{figure}[t]
\begin{center}
\includegraphics[width=\columnwidth]{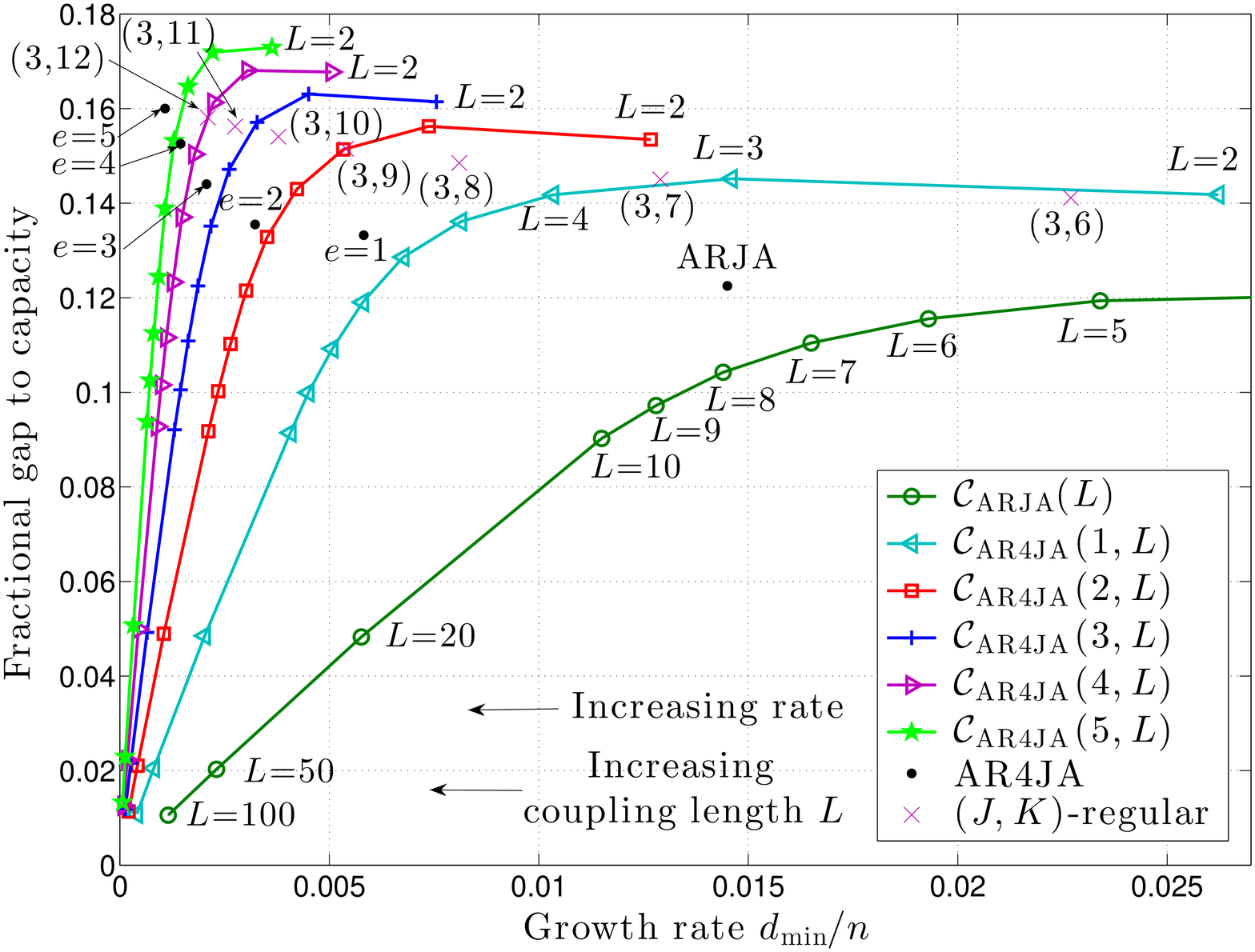}
\end{center}
\caption{Minimum distance growth rate vs. the fractional gap to capacity for the $\mathcal{C}_\mathrm{AR4JA}(e,L)$ SC-LDPC-BC ensembles, the underlying AR4JA LDPC-BC ensembles, and several $(J,K)$-regular LDPC-BC ensembles.}\label{fig:distvspropgap}
\end{figure}

Fig.~\ref{fig:distvspropgap} shows the minimum distance growth rates against the fractional gap to capacity $(\varepsilon_{sh}-\varepsilon^*)/\varepsilon_{sh}$ for the $\mathcal{C}_\mathrm{AR4JA}(e,L)$ SC-LDPC-BC ensembles with coupling lengths $L=2,\ldots,10,20,50,100$, the underlying AR4JA LDPC-BC ensembles, and several $(J,K)$-regular LDPC-BC ensembles. The trade-off we observe effectively allows a code designer to `tune' between distance growth rate and threshold by choosing the parameters $e$ and $L$. We observe that, in particular, intermediate values of $L$ provide thresholds with a small gap to capacity while maintaining a reasonable distance growth rate with only a small loss in code rate. The complexity of the $\mathcal{C}_\mathrm{AR4JA}(e,L)$ ensembles (measured by average variable and check node degrees) increases slowly with $L$ and approaches that of the underlying AR4JA LDPC-BC ensemble for a given extension parameter $e$. Further, as $L$ becomes sufficiently large for the scaled growth rates to converge, we observe that the gaps to capacity are approximately proportional to $L$ for all of the $\mathcal{C}_\mathrm{AR4JA}(e,L)$ ensembles. For example, we obtain about a $10\%$ gap to capacity by terminating after $L=9$ time instants; a $5\%$ gap after $L=20$ time instants; a $2\%$ gap after $L=50$ time instants; and a $1\%$ gap after $L=100$ time instants. Finally, choosing the extension parameter $e$ allows additional flexibility, where a larger $e$ gives a higher code rate but a lower distance growth rate and greater complexity. \hfill $\Box$\medskip

\subsection{Thresholds for the AWGNC}\label{sec:awgn}
In this section, we perform an AWGNC threshold analysis of protograph-based SC-LDPC-BC ensembles and show that the dramatic threshold improvement obtained by terminating SC-LDPC-CC also extends to the AWGNC. Since exact DE is far more complex for the AWGNC than for the BEC, it is only feasible for simple protographs, so here we make use of the reciprocal channel approximation (RCA) technique introduced in \cite{chu00}, which has been successfully applied to the analysis of protograph-based ensembles in \cite{ddja09}. With this approach, the calculation of approximate AWGNC thresholds for a variety of regular and irregular protographs becomes feasible with reasonable accuracy.


\par
\begin{figure}[t]
\begin{center}
\includegraphics[width=\columnwidth]{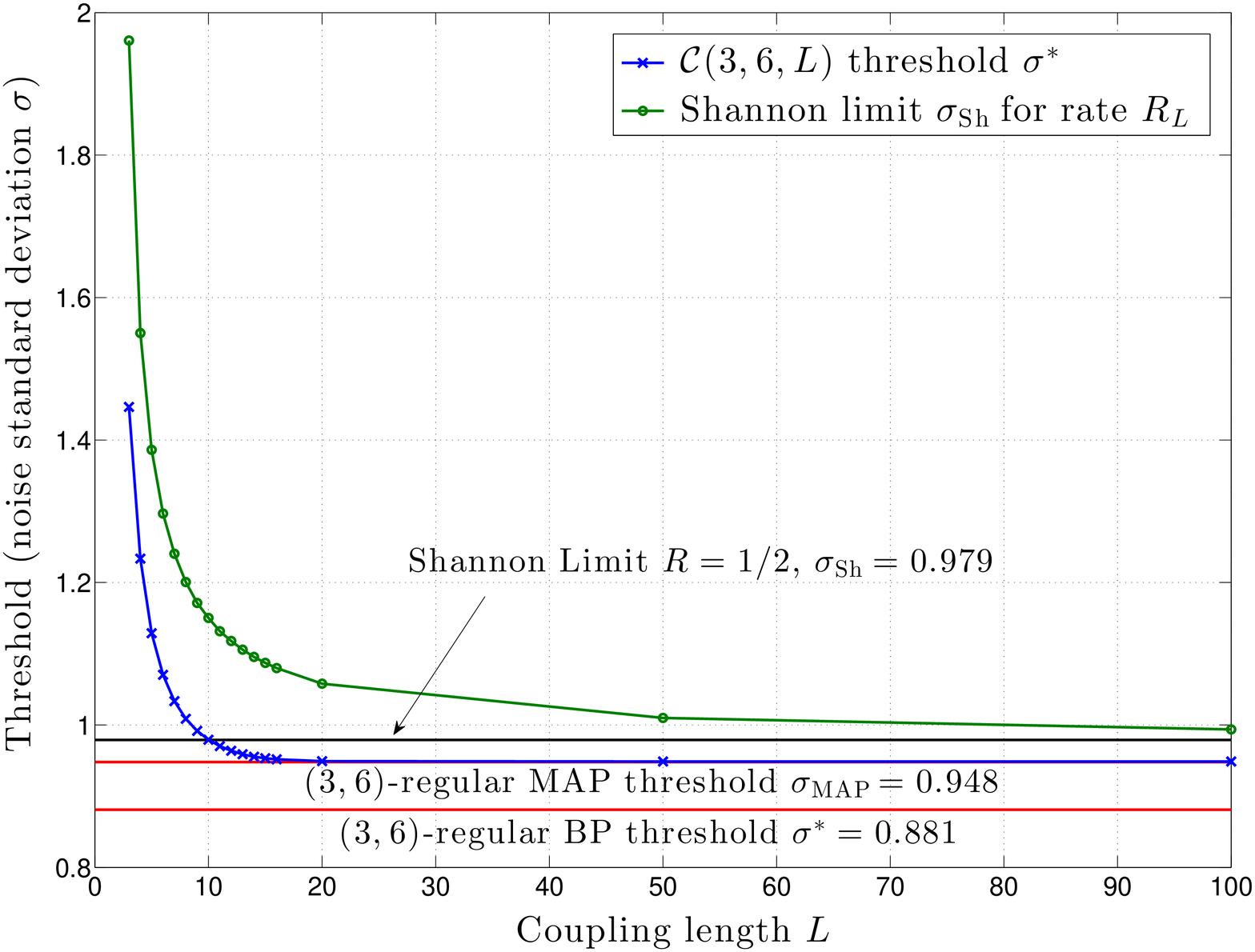}
\end{center}
\caption{AWGNC BP thresholds in terms of the noise standard deviation $\sigma$ for $\mathcal{C}(3,6,L)$ SC-LDPC-BC ensembles with design rate $R_L=(L+2)/2L$ and the corresponding Shannon limit for rate $R_L$. Also shown for comparison are the BP and MAP decoding thresholds for the underlying $(3,6)$-regular LDPC-BC ensemble, $\sigma^*=0.881$ and $\sigma_{\mathrm{MAP}}=0.948$, respectively, and the Shannon limit for $R=1/2$ codes, $\sigma_{\mathrm{Sh}}=0.979$.}\label{fig:36thresAWGNC}
\end{figure}

\medskip\noindent{\bf Example \ref{ex:36proto} (cont.).} In Fig.~\ref{fig:36thresAWGNC}, we plot the AWGNC BP thresholds (in terms of the noise standard derivation $\sigma$) obtained using the RCA technique for several $\mathcal{C}(3,6,L)$ SC-LDPC-BC ensembles along with the Shannon limit for the given design rate $R_L=(L-2)/2L$. We observe the same behavior as demonstrated for the BEC in Fig.~\ref{fig:r12thres}: we find that the threshold  decreases monotonically with increasing rate, but the gap to capacity for the given rate also decreases. For example, the threshold values are equal to $\sigma^* = 1.446$ for $L = 3$ and $\sigma^* = 0.9638$ for $L = 10$. As $L$ is further increased, the thresholds saturate to a constant value $\sigma^*=0.948$ and do not further decay as $L\rightarrow\infty$ and $R_L\rightarrow 1/2$. We observe that $\sigma^*=0.948$ which is equal to the MAP decoding threshold of the underlying $(3,6)$-regular LDPC-BC ensemble, is much closer to the Shannon limit $\sigma_\mathrm{Sh}=0.979$ than the BP threshold $\sigma^*=0.881$ of the $(3,6)$-regular LDPC-BC ensemble. \hfill $\Box$\medskip

\begin{figure}
\begin{center}
\small
\includegraphics[width=\columnwidth]{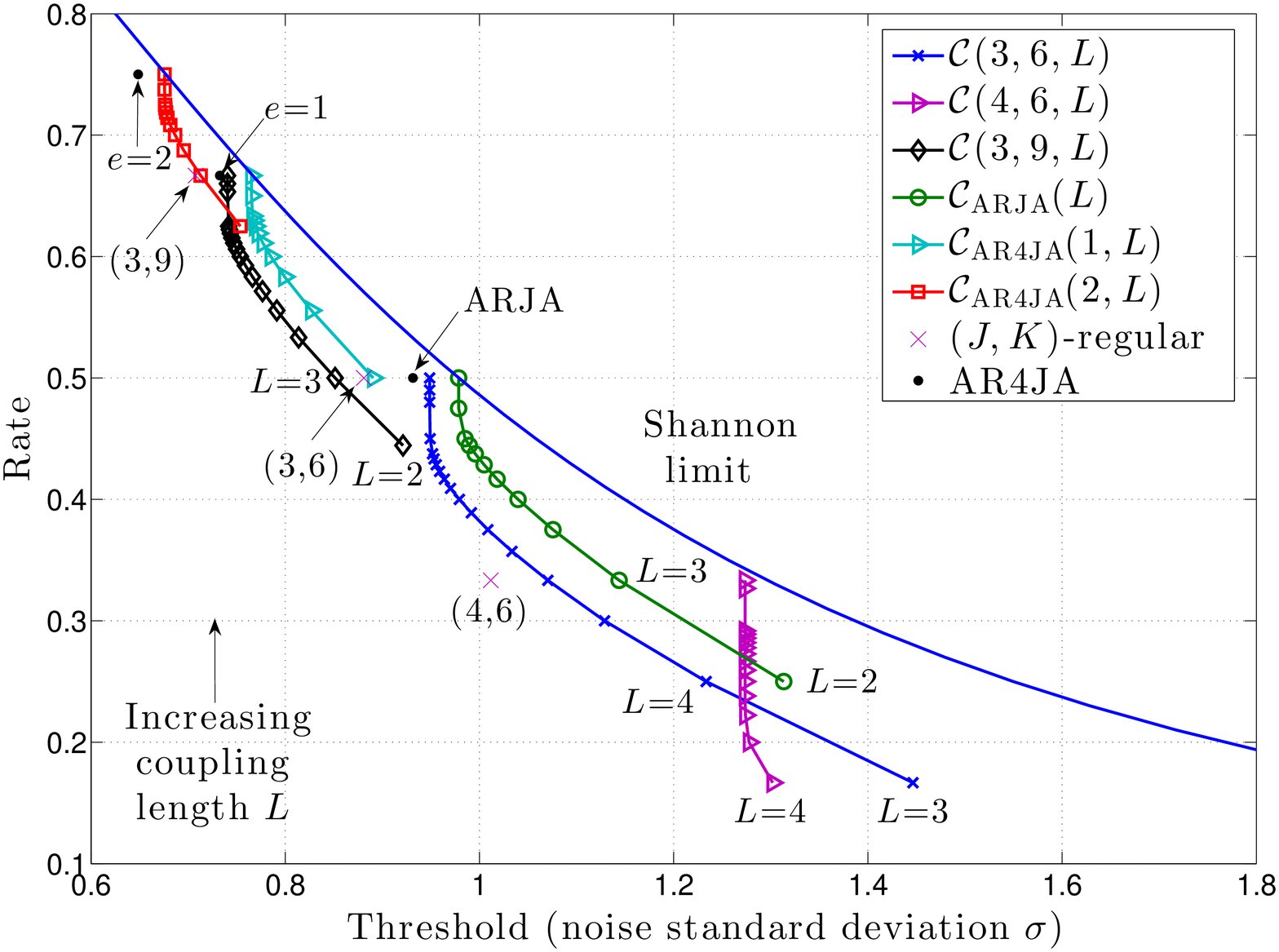}\\(a)\vspace{2mm}

\includegraphics[width=\columnwidth]{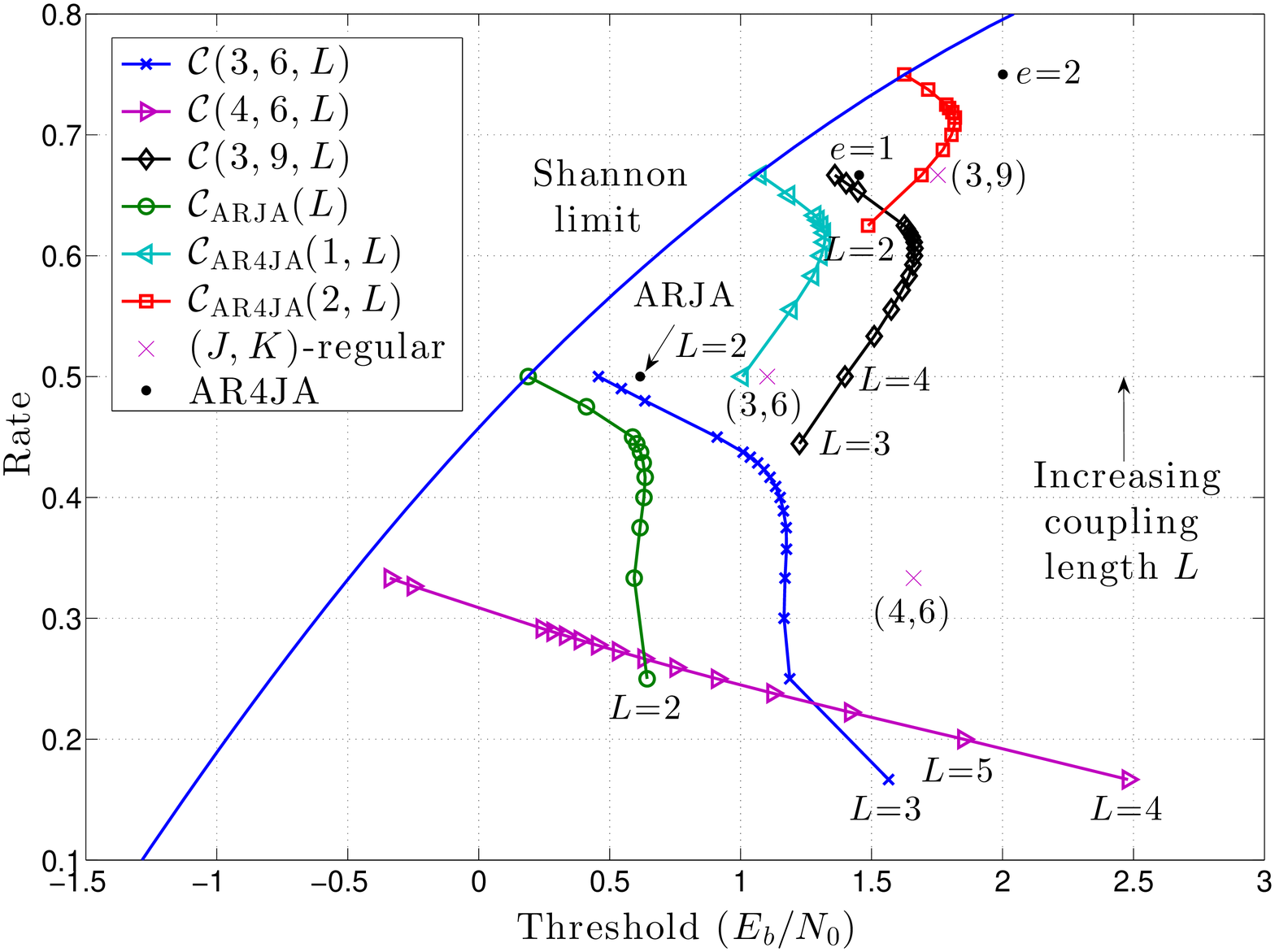}\\(b)

\caption{AWGNC BP thresholds in terms of (a) standard deviation $\sigma$ and (b) signal-to-noise ratio $E_b/N_0$ (dB) for several families of SC-LDPC-BC ensembles with different coupling lengths $L$.} \label{fig:ThresEbN0} 
\end{center}
\end{figure}

This behavior, which can be observed for all of the regular and irregular ensembles considered, is similar to the corresponding results for the BEC, presented in Section~\ref{sec:bec}. In Fig.~\ref{fig:ThresEbN0}, we display calculated  thresholds for the AWGNC for a variety of $\mathcal{C}(J,K,L)$, $\mathcal{C}_\mathrm{ARJA}(L)$, and $\mathcal{C}_\mathrm{AR4JA}(e,L)$ ensembles along with the thresholds of the underlying LDPC-BC ensembles. Fig.~\ref{fig:ThresEbN0}(a) plots the thresholds in terms of the standard deviation $\sigma$ against the ensemble design rate. We observe that  as $L$ increases, the design rate
 increases and the threshold decreases monotonically; however, as $L$ becomes sufficiently large, the thresholds saturate to the MAP thresholds of the underlying LDPC-BC ensembles. Further, they are close to the Shannon limit and, importantly, they do not decrease further as $L\rightarrow\infty$ and the design rate and degree distribution approach those of the underlying LDPC-BC ensembles. 

The same thresholds are depicted in Fig.~\ref{fig:ThresEbN0}(b) in terms of the SNR $E_b/N_0$. Since $E_b/N_0$ takes into account the code rate overhead, the ensembles with lower rate have a larger noise variance, and the  monotonic behavior of the thresholds noted in Fig.~\ref{fig:ThresEbN0}(a) is  no longer visible. In both plots, however,  we see that the gap to capacity decreases with increasing $L$. Consequently, in a similar fashion to the BEC, varying the coupling length $L$ results in SC-LDPC-BC ensembles with different design rates and a trade-off between iterative decoding threshold and minimum distance growth rate.

To conclude this section, we examine the AWGNC BP thresholds for various SC-LDPC-BC ensembles with design rates approaching $R_\infty = 1/2$ and varying graph densities. In addition, we use simulation to examine the finite length performance of SC-LDPC-BC ensembles on the AWGNC and demonstrate that the excellent performance promised by the asymptotic results also translates into improved decoding performance for finite code lengths.

\par
\begin{figure}[t]
\begin{center}\small
\includegraphics[width=\columnwidth]{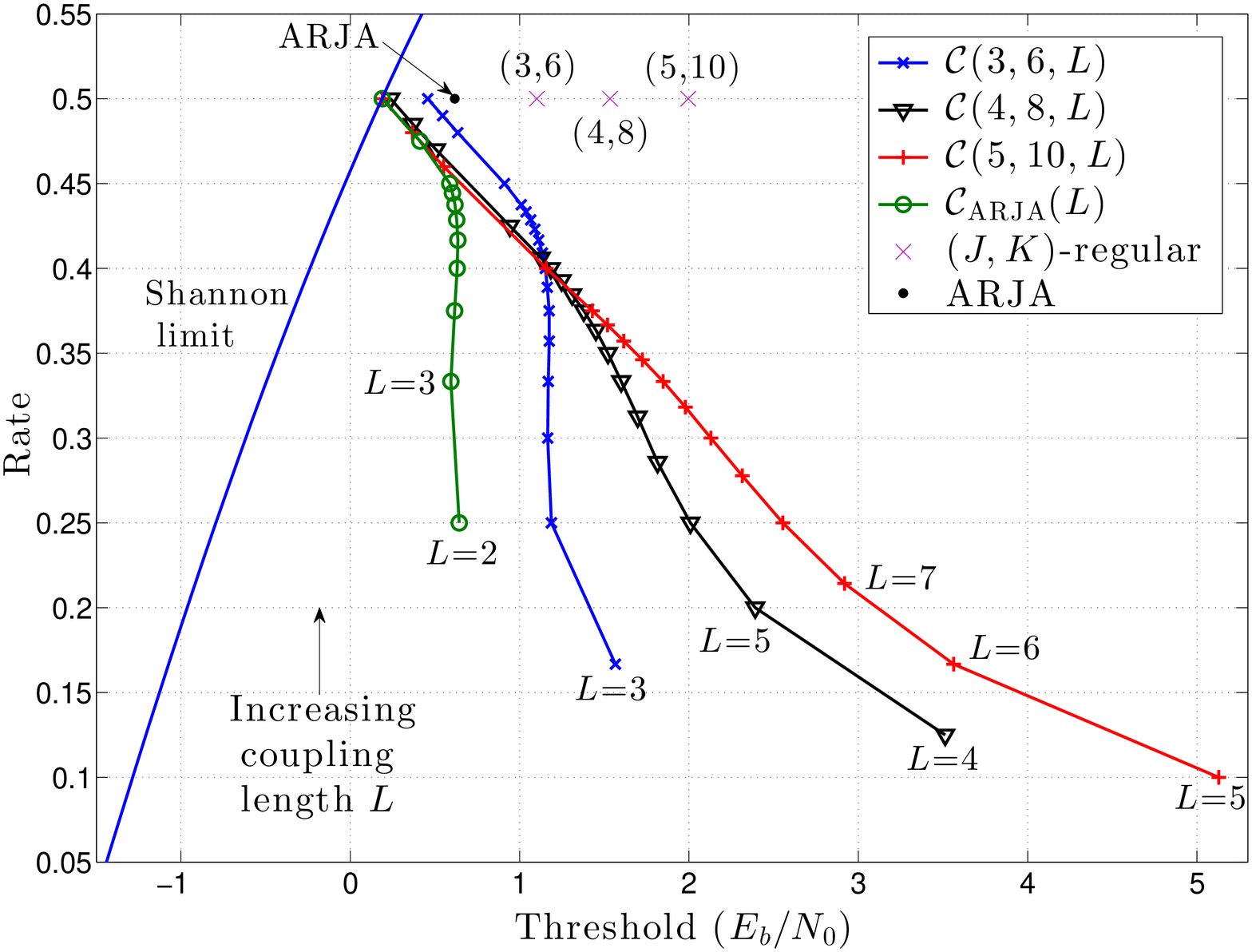}\\(a)\vspace{2mm}

\includegraphics[width=\columnwidth]{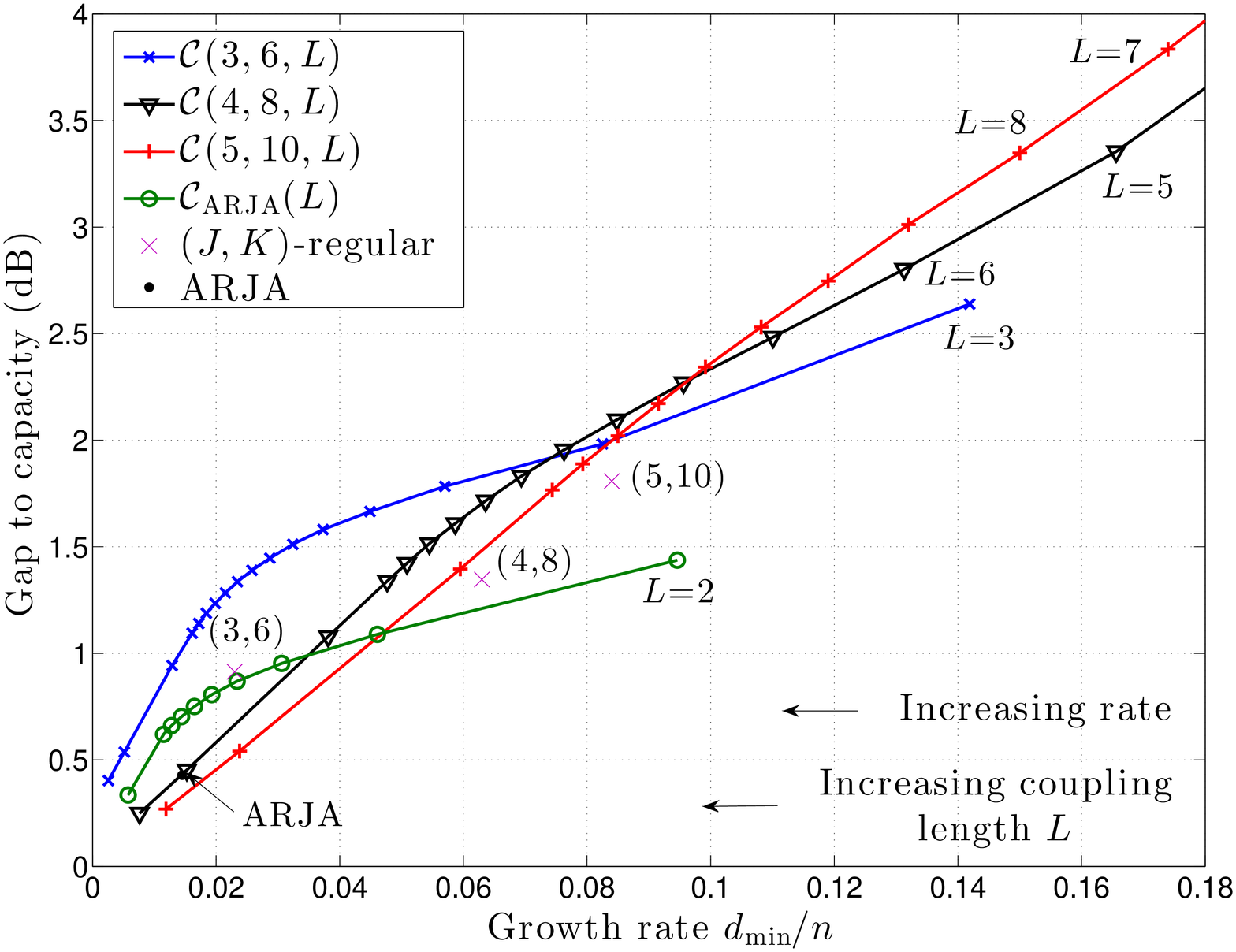}\\(b)
\end{center}
\caption{Comparison of the $\mathcal{C}(J,2J,L)$ SC-LDPC-BC ensembles, the $\mathcal{C}_\mathrm{ARJA}(L)$ SC-LDPC-BC ensembles, and the underlying LDPC-BC ensembles: (a) AWGNC thresholds in terms of SNR $E_b/N_0$ (dB), and (b) minimum distance growth rate vs. threshold gap to capacity.}\label{fig:awgnthreshdist}
\end{figure}

\medskip\noindent{\bf Example \ref{ex:r12} (cont.).}  We consider once more several $\mathcal{C}(J,2J,L)$ ensembles and the $\mathcal{C}_\mathrm{ARJA}(L)$ ensembles in Fig.~\ref{fig:awgnthreshdist}(a). We observe that, as $L$ increases, the ensemble design rate of the SC-LDPC-BC ensembles increases (approaching $R=1/2$ asymptotically) and the thresholds improve, nearing the Shannon limit for large $L$. Further, we note that the $(J,2J)$-regular LDPC-BC ensemble thresholds worsen as we increase $J$, and we see that this is also the case for the $\mathcal{C}(J,2J,L)$ ensembles on the AWGNC for small coupling lengths $L$, as was previously noted for the BEC. Thus, for $(J,2J)$-regular LDPC-BC ensembles and $\mathcal{C}(J,2J,L)$ ensembles with small $L$, thresholds worsen and distance growth rates improve by increasing $J$ (and hence iterative decoding complexity). However, as $L$ increases, the behavior of the $\mathcal{C}(J,2J,L)$ ensembles changes and their thresholds saturate to a value numerically indistinguishable from the MAP decoding threshold of the underlying $(J,2J)$-regular LDPC-BC ensemble and this value approaches  the Shannon limit as we increase $J$. This indicates that, for large $L$, both the distance growth rates and the thresholds improve with increasing complexity, and we  expect this trend to continue as we further increase the variable node degree $J$, although the improvement will diminish with increasing $J$. As a final observation, we note that for all achievable rates the $\mathcal{C}_\mathrm{ARJA}(L)$ ensembles have better thresholds than the $\mathcal{C}(3,6,L)$ ensembles. This is expected, since the ARJA LDPC-BC ensemble has been designed to have a good iterative decoding threshold. However, we also observe that, for large $L$, the $\mathcal{C}(4,8,L)$ and $\mathcal{C}(5,10,L)$ ensembles have comparable thresholds to the $\mathcal{C}_\mathrm{ARJA}(L)$ ensembles, demonstrating the benefit that derives from the spatially coupled convolutional structure, \emph{i.e.}, we obtain near capacity threshold performance with an almost regular code graph.

Fig.~\ref{fig:awgnthreshdist}(b) plots the minimum distance growth rates against the threshold gap to capacity (the difference between the AWGNC threshold (in terms of $E_b/N_0$) of an ensemble and capacity for the ensemble design rate) for the $\mathcal{C}(J,2J,L)$ ensembles with coupling lengths $L=w+1,\ldots,16,20,50,100$, the $\mathcal{C}_\mathrm{ARJA}(L)$ ensembles with $L=2,\ldots,10$, and the underlying $(J,2J)$-regular and ARJA LDPC-BC ensembles.  We observe that, in particular, intermediate values of $L$ provide thresholds with a small gap to capacity while maintaining linear minimum distance growth with only a slight loss in code rate. We also note that, for a fixed gap to capacity close to zero, the largest minimum distance growth rate is obtained by choosing the $\mathcal{C}(J,2J,L)$ ensemble with the largest $J$, and that the $\mathcal{C}_\mathrm{ARJA}(L)$ ensembles falls in between the  $\mathcal{C}(3,6,L)$ and $\mathcal{C}(4,8,L)$ ensembles. (In this region, with the gap to capacity close to zero, the rates of all the ensembles are approximately equal and close to $1/2$.) For larger fixed gaps to capacity, we see that the order changes and that the reverse order holds for large gaps to capacity. 

\begin{figure}[t]
\centering
\includegraphics[width=\columnwidth]{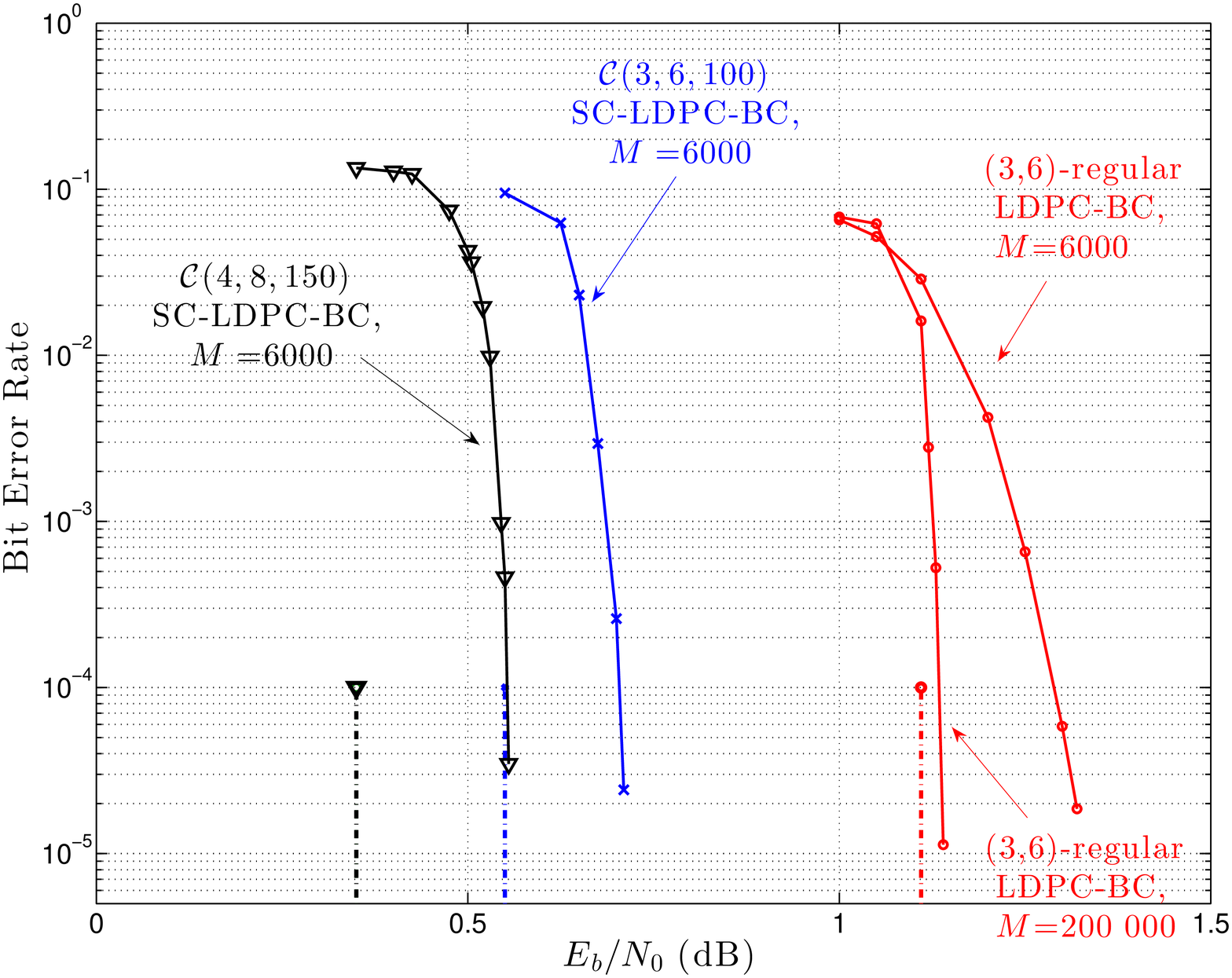}
\caption{AWGNC decoding performance (solid lines) and BP decoding thresholds (dashed lines) of $\mathcal{C}(3,6,100)$ and $\mathcal{C}(4,8,150)$ SC-LDPC-BCs with lifting factor $M=6000$ and rate $R=0.49$. For comparison, the performance of $(3,6)$-regular LDPC-BCs with $M=6000$ and $M=200 000$ are also shown along with the associated BP decoding threshold. (The BP thresholds of $(4,8)$- and $(5,10)$-regular LDPC-BCs are $1.61$dB and $2.04$dB, respectively.)}\label{fig:threshperformance}
\end{figure}

Now consider choosing $L$ such that the ensemble design rate is $R=0.49$ and the BP threshold values of the $\mathcal{C}(J,2J,L)$ ensembles improve with $J$. The thresholds of the $\mathcal{C}(3,6,100)$ and $\mathcal{C}(4,8,150)$ ensembles are shown in Fig.~\ref{fig:threshperformance}, along with the simulated performance of randomly chosen codes from the underlying ensembles with permutation matrix size $M=6000$. Also shown, for comparison, is the BP threshold of the $(3,6)$-regular LDPC-BC with design rate $R=0.5$ along with the simulated decoding performance of two randomly chosen codes from the ensemble with permutation matrix sizes $M=6000$ and $M=200,000$. A standard LDPC-BC decoder employing the BP decoding algorithm was used in each case. For the SC-LDPC-BCs with $M=6000$, we observe that the waterfall performance is within $0.2$dB of  the threshold, and we expect the gap to decrease for larger permutation matrix sizes $M$. By choosing a larger $L$, the rate increases (approaching $1/2$), and the thresholds and corresponding waterfall performance of codes chosen from these ensembles will improve slightly. Note, in particular, that the SC-LDPC-BCs are operating far beyond the threshold of the $(3,6)$-regular LDPC-BC ensemble, and as  the graph density $J$ increases this improvement will become more pronounced, since the thresholds and corresponding waterfall performance of the SC-LDPC-BCs gets better whereas the thresholds and performance of the LDPC-BCs will become worse.\hfill $\Box$\medskip

\section{Free distance growth rates of SC-LDPC-CC ensembles}\label{sec:growth}


In the previous section, we saw that the BP decoding thresholds of SC-LDPC-BC ensembles tend toward the MAP thresholds of the underlying LDPC-BC ensembles on both the BEC and AWGNC with increasing coupling length $L$, while the minimum distance growth rates $\delta_\mathrm{min}^{(L)}$ tend to zero as $L \rightarrow \infty$. In a practical code design, for a given finite block length $n = M Lb_v $, a careful choice of the parameters $M$ and $L$ becomes necessary to achieve the best performance. The minimum distance growth rate $\delta_\mathrm{min}^{(L)}$ provides a useful measure for comparison of the distance properties of SC-LDPC-BC ensembles; however, based on their spatially coupled (convolutional) structure, it is clear that the potential strength of SC-LDPC-BC ensembles for large $L$ scales with the constraint length $\nu=M (w+1)b_v $, which increases with $M$ but is independent of $L$. Consequently, 
the free distance growth rate $\delta_\mathrm{free}$ of the closely related SC-LDPC-CC ensemble, which is independent of $L$, is a more appropriate measure of the performance of SC-LDPC-BC ensembles than their minimum distance growth rates $\delta_\mathrm{min}^{(L)}$.

This fact is supported by the excellent decoding performance of a continuous sliding window decoder \cite{lscz10,lpf11,ips+12}, which only passes messages across a window of fixed size $W$, typically a small multiple of the constraint length $\nu$ (independent of $L$), as opposed to passing messages directly across the entire length of the graph (which grows with $L$), like a standard LDPC-BC BP decoder. Provided that $W$ is not chosen to be too small, the strength of the SC-LDPC-BC is contained within the window and there is no perceivable loss in performance compared to a standard decoder \cite{lpf11,ips+12}. Moreover, this strategy is of particular practical importance since it allows one to fully exploit the localized structure of SC-LDPC-BC ensembles in terms of minimizing decoding latency and memory requirements.

In the remainder of this section, we investigate the connection between the minimum distance of SC-LDPC-BC ensembles and the free distance of the closely related SC-LDPC-CC ensembles.

\begin{definition}\vspace{-2mm}\emph{
The minimum \emph{free distance} of a convolutional code, denoted by $d_\mathrm{free}$, is defined as the minimum Hamming distance between any two distinct code sequences in the code $\mathbf{x}_{[0,\infty]}$ and $\mathbf{y}_{[0,\infty]}$. Since convolutional codes are linear, this condition simplifies to 
\begin{equation*}
 d_\mathrm{free} = \min_{\mathbf{x}_{[0,\infty]} \neq \mathbf{0}} w(\mathbf{x}_{[0,\infty]}),
\end{equation*}
where $w(\cdot)$ denotes the Hamming weight of the argument, i.e., $d_\mathrm{free}$ is the weight of the minimum Hamming weight nonzero code sequence.}
\end{definition}

In order to obtain our result, we introduce a sub-ensemble of the SC-LDPC-CC ensemble given in Definition \ref{def:ldpccc} where each member is \emph{periodically time-varying}.

\begin{definition}\label{def:period}\vspace{-2mm}
\emph{An ensemble of \emph{periodically time-varying} protograph-based SC-LDPC-CCs with coupling width $w$, design rate $R=1-b_c/b_v$, \emph{constraint length} $\nu=M(w+1)b_v$, and period $T$  is obtained as the collection of all $M$-fold graph covers of a convolutional protograph where the permutation applied to edge $l$ of variable node $v_y$, $l \in \{1,\ldots, \partial(v_y)\}$, $y \in \{0,1,\ldots,b_v-1\}$, at time $t$ is also applied to edge $l$ at times $\{t+kT|k\in\mathbb{Z}\backslash0\}$ and $T$ is the smallest natural number for which this condition holds.}
\end{definition}

To avoid confusion with the notation, we will refer to the periodically time-varying SC-LDPC-CC sub-ensembles of the $\mathcal{C}(J,K)$, $\mathcal{C}_\mathrm{ARJA}$, and $\mathcal{C}_\mathrm{AR4JA}(e)$ SC-LDPC-CC ensembles with period $T$ as $\mathcal{T}(J,K,T)$, $\mathcal{T}_\mathrm{ARJA}(T)$, and $\mathcal{T}_\mathrm{AR4JA}(e,T)$, respectively. It is known that the average free distance of an ensemble of periodically time-varying protograph-based SC-LDPC-CCs with period $T$ constructed as described in Definition \ref{def:period} can be bounded below by the average minimum distance of the associated ensemble of TB-SC-LDPC-BCs derived from the base matrix $\mathbf{B}^{\mathrm{tb}}_{[0,L-1]}$ with coupling length $L=T$ \cite{tzc10,mpc13}. Here, we show that the average free distance of this ensemble can also be bounded above by the average minimum distance of the terminated SC-LDPC-BC ensemble derived from the base matrix $\mathbf{B}_{[0,L-1]}$ with $L=T$. 

\begin{theorem}\emph{
Consider a rate $R=1-b_c/b_v$ periodically
time-varying SC-LDPC-CC ensemble with coupling width $w$, constraint length
$\nu=M(w+1)b_v$, and period $T$ derived from a convolutional protograph with base matrix $\mathbf{B}_{[-\infty,\infty]}$. Let $\overline{d}^{(L)}_\mathrm{min}$ be
the average minimum distance of the associated SC-LDPC-BC ensemble with coupling length
$L$ and block length $n=ML b_v$ derived from the terminated convolutional protograph with base matrix $\mathbf{B}_{[0,L-1]}$. Then the ensemble average free distance $\overline{d}_\mathrm{free}^{(T)}$ of the SC-LDPC-CC ensemble is bounded above by $\overline{d}_\mathrm{min}^{(L)}$ for coupling length $L=T$, i.e.,
\begin{equation}\label{dfreebound}
    \overline{d}_\mathrm{free}^{(T)} \leq \overline{d}^{(T)}_\mathrm{min}.
\end{equation}}
\end{theorem}
\emph{Proof}. There is a one-to-one relationship between members of the periodically time-varying SC-LDPC-CC ensemble and members of the associated SC-LDPC-BC ensemble with coupling length $L=T$. For any such pair of codes, every codeword $\mathbf{x}_{[0,M L b_v-1]}=[\begin{array}{cccc} x_0 & x_1 & \cdots & x_{M L b_v-1}\end{array}]$ in the terminated code be viewed as a codeword $\mathbf{x}_{[0,\infty]}=[\begin{array}{ccccccc} x_0 & x_1 & \cdots & x_{L N b_v-1}&0&0&\cdots\end{array}]$ in the unterminated code. It follows that the free distance $d_\mathrm{free}^{(T)}$ of the unterminated code can not be larger than the minimum distance $d_\mathrm{min}^{(T)}$ of the terminated code. 
The ensemble average result $\overline{d}_\mathrm{free}^{(T)} \leq \overline{d}^{(T)}_\mathrm{min}$ then follows directly. \hfill $\Box$\medskip

\noindent Since there is no danger of ambiguity, we will henceforth drop the overline notation when discussing ensemble average distances. 

For SC-LDPC-CCs, conventionally defined as the null space of a sparse parity-check matrix $\mathbf{H}_{[0,\infty]}$, it is natural to define the free distance growth rate with respect to the constraint length $\nu$, i.e., as the ratio of the free distance $d_\mathrm{free}$ to the constraint length $\nu$. 
By bounding $d_\mathrm{free}^{(T)}$ using (\ref{dfreebound}), we obtain an upper bound on the free distance growth rate as
\begin{equation}\label{ub}
\delta_\mathrm{free}^{(T)}=\frac{d_\mathrm{free}^{(T)}}{\nu} \leq \frac{{\delta}_\mathrm{min}^{(T)}T}{(w+1)},
\end{equation}
where ${\delta}_\mathrm{min}^{(T)} = {d_\mathrm{min}^{(T)}}/{n}={d_\mathrm{min}^{(T)}}/{(MT  b_v)}$ is the minimum distance growth rate of the  SC-LDPC-BC ensemble with coupling length $L=T$ and base matrix $\mathbf{B}_{[0,T-1]}$.\footnote{The free distance growth rate $\delta_\mathrm{free}^{(T)}$ that we bound from above using (\ref{ub}) is, by definition, an existence-type lower bound on the free distance typical of most members of the ensemble, i.e., with high probability a randomly chosen code from the ensemble has free distance at least as large as ${\delta}_\mathrm{free}^{(T)}\nu$ as $\nu\rightarrow \infty$.}  Similarly, it was shown in \cite{mpc13} that
\begin{equation}\label{lb}
\delta_\mathrm{free}^{(T)}\geq \frac{\check{\delta}_\mathrm{min}^{(T)}T}{(w+1)},
\end{equation}
where $\check{\delta}_\mathrm{min}^{(T)}$ is the minimum distance growth rate of the TB-SC-LDPC-BC ensemble with tail-biting coupling length $L=T$ and base matrix $\mathbf{B}^{\mathrm{tb}}_{[0,L-1]}$.

\par
\begin{figure}[h]
\begin{center}
\includegraphics[width=\columnwidth]{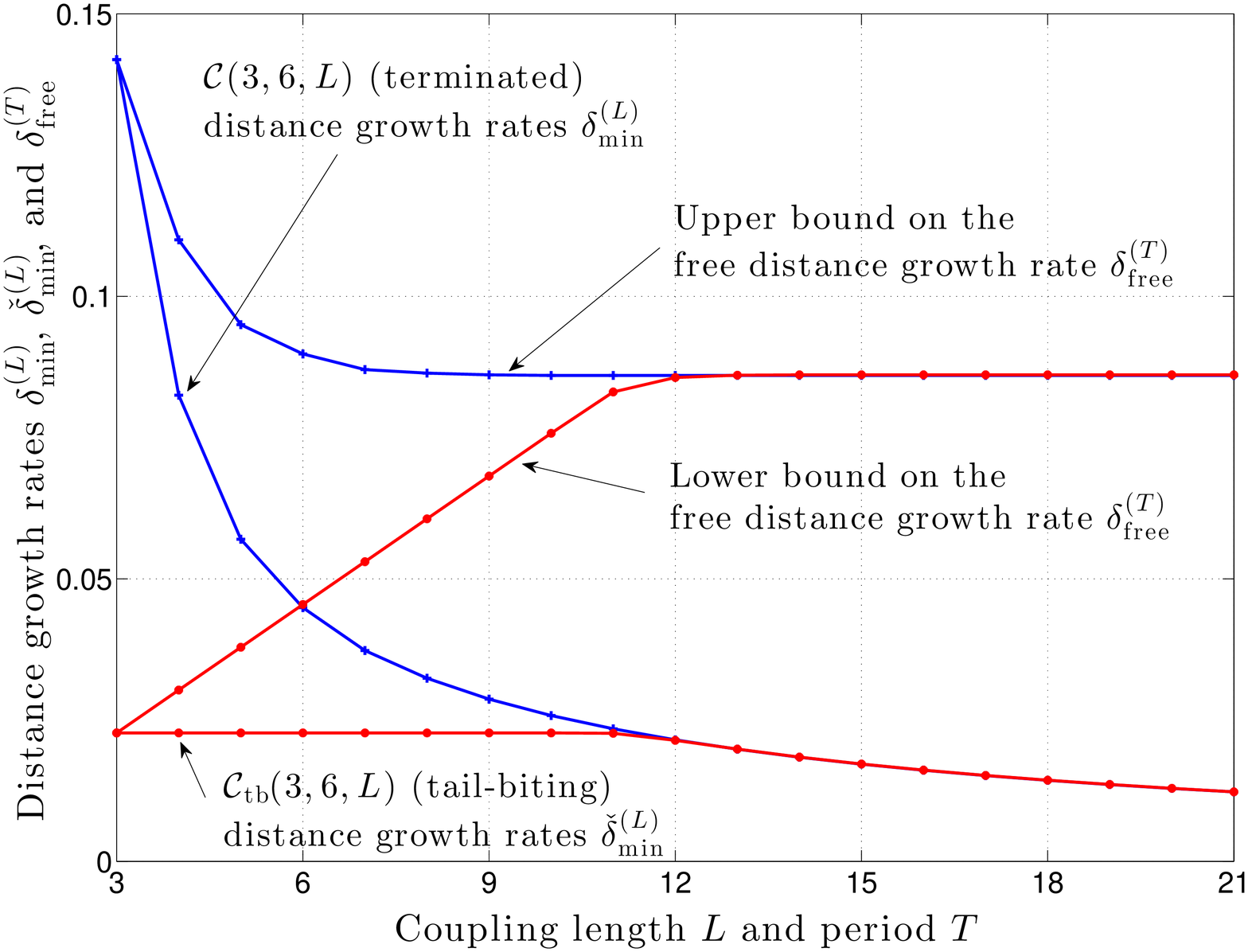}
\end{center}
\caption{Minimum distance growth rates of the $\mathcal{C}(3,6,L)$ (terminated) and $\mathcal{C}_\mathrm{tb}(3,6,L)$ (tail-biting) SC-LDPC-BC ensembles with upper and lower bounds on the free distance growth rate of the associated periodically time-varying SC-LDPC-CC ensembles $\mathcal{T}(3,6,L)$.}\label{fig:36bounds}\end{figure}

\medskip\noindent{\bf Example \ref{ex:36proto} (cont.).} As an example, consider the $\mathcal{C}(3,6,L)$ SC-LDPC-BC ensembles. Using (\ref{ub}), we calculate the upper bound on the free distance growth rate of the periodically time-varying SC-LDPC-CC ensemble $\mathcal{T}(3,6,T)$, with design rate $R=1/2$, as $\delta_\mathrm{free}^{(T)} \leq {\delta}^{(T)}_\mathrm{min}T/3$  for coupling lengths $L=T\geq 3$. Fig.~\ref{fig:36bounds} displays the minimum distance growth rates ${\delta}^{(L)}_\mathrm{min}$ of the $\mathcal{C}(3,6,L)$ ensembles defined by $\mathbf{B}_{[0,L-1]}$ for $L=3,4,\ldots,21$ and the associated upper bounds on the free distance growth rate $\delta_\mathrm{free}^{(T)} \leq {\delta}^{(T)}_\mathrm{min}T/3$ for $L=T$. Also shown are the minimum distance growth rates $\check{\delta}^{(L)}_\mathrm{min}$ of the  $\mathcal{C}_\mathrm{tb}(3,6,L)$ TB-SC-LDPC-BC ensembles,    defined by base matrix $\mathbf{B}^{\mathrm{tb}}_{[0,L-1]}$ for $L=3,4,\ldots,21$, and the associated lower bounds, calculated using (\ref{lb}),  on the free distance growth rate $\delta_\mathrm{free}^{(T)} \geq \check{\delta}^{(T)}_\mathrm{min}T/3$ for $L=T$.

We observe that the $\mathcal{C}_\mathrm{tb}(3,6,L)$ minimum distance growth rates $\check{\delta}_\mathrm{min}^{(L)}$ remain constant for $L=3, \ldots, 11$ and then start to decrease as the coupling length $L$ grows, tending to zero as $L$ tends to infinity. Correspondingly, as $L$ exceeds $11$, the lower bound on $\delta_\mathrm{free}^{(T)}$ levels off at $\delta_\mathrm{free}^{(T)} \geq 0.086$. As discussed in Section~\ref{sec:weight}, the $\mathcal{C}(3,6,L)$ minimum distance growth rates ${\delta}_\mathrm{min}^{(L)}$ are large for small values of $L$ (where the rate loss is larger) and decrease monotonically to zero as $L\rightarrow \infty$. Using (\ref{ub}) to obtain an upper bound on the free distance growth rate we observe that, for $T\geq12$, the upper and lower bounds on $\delta_\mathrm{free}^{(T)}$ coincide, indicating that, for these values of the period $T$, $\delta_\mathrm{free}^{(T)}=0.086$, significantly larger than the underlying  $(3,6)$-regular LDPC-BC ensemble minimum distance growth rate $\delta_\mathrm{min}=0.023$. This leveling-off phenomenon occurs as a result of the fact that the minimum weight codeword in a typical member of the SC-LDPC-CC ensemble also appears as a codeword in a typical member of the SC-LDPC-BC ensemble once $L$ exceeds $11$. In addition, we note that, at the point where the bounds coincide, the minimum distance growth rates for both the terminated and tail-biting ensembles coincide. (Recall that the bounds diverge for smaller values of $L$ since the $\mathcal{C}_\mathrm{tb}(3,6,L)$ ensembles have rate $1/2$ for all $L$, whereas the rate of the $\mathcal{C}(3,6,L)$ ensembles is a function $L$ given by (\ref{termrate}).) \hfill $\Box$\medskip

\begin{figure}[t]
\begin{center}
\includegraphics[width=\columnwidth]{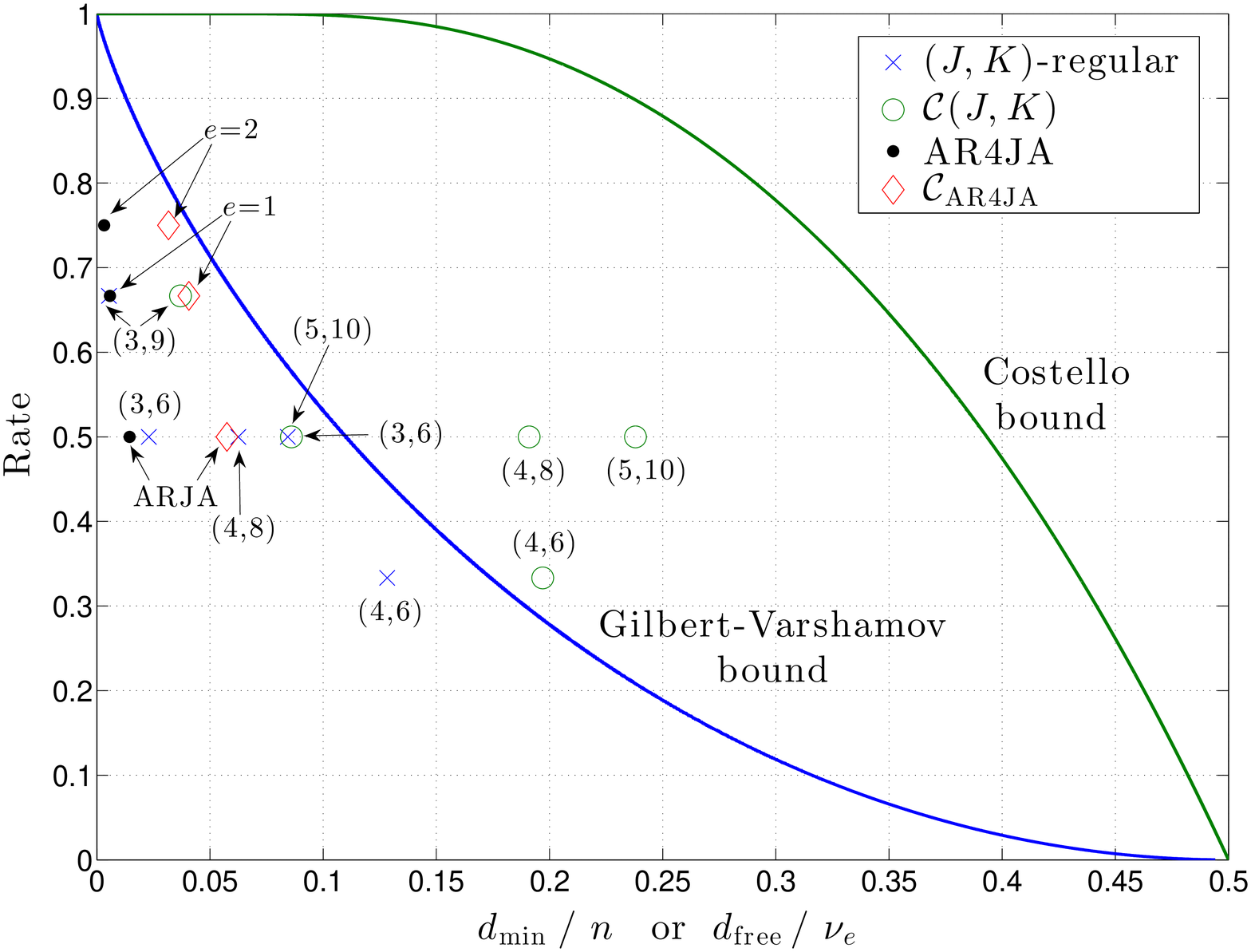}
\end{center}
\caption{Asymptotic free distance growth rates for some $\mathcal{C}(J,K)$ and $\mathcal{C}_\mathrm{AR4JA}(e)$ SC-LDPC-CC ensembles.} \label{fig:dfreegrowth} \end{figure}

Numerically, it becomes problematic to evaluate $\delta_{min}^{(L)}$  and $\check{\delta}_{min}^{(L)}$ for large values of $L$, but the leveling-off effect noted in Fig.~\ref{fig:36bounds}, which also occurs in all the other cases we have examined, strongly suggests that the free distance growth rate $\delta_{free}^{(T)}$ remains constant once $T$ increases beyond a certain value. This is due to the fact that, for a fixed constraint length, further increases in the period cannot result in convolutional ensembles with larger free distances. We set $\delta_{free} \triangleq  \max_{T} \delta_{free}^{(T)}$, and the leveling-off numerical results obtained for the $\mathcal{T}(3,6,T)$ ensembles suggests that the free distance growth rate $\delta_{free}$ of the associated (non-periodic) $\mathcal{C}(3,6)$ SC-LDPC-CC ensemble converges to $0.086$. Lower bounds on free distance growth rates were calculated for a wide variety of $(J,K)$-regular and irregular protograph-based SC-LDPC-CC ensembles in \cite{mpc13} and, using the technique described here, we can form upper bounds on the free distance growth rates that coincide numerically with the lower bounds for sufficiently large $T$, resulting in exact free distance growth rates. Growth rates for a variety of $\mathcal{C}(J,K)$ and $\mathcal{C}_\mathrm{AR4JA}(e)$ SC-LDPC-CC ensembles are plotted in Fig.~\ref{fig:dfreegrowth}, along with the minimum distance growth rates of the underlying LDPC-BC ensembles. Also shown are the Gilbert-Varshamov and Costello \cite{cos74} lower bounds on the growth rates of general ensembles of random block and convolutional codes, respectively.\footnote{The constraint length $\nu$ that we define in this paper (see Definition \ref{def:ldpccc}) is often referred to as the \emph{decoding constraint length}. In order to facilitate comparison to the Costello bound, rather than using \eqref{ub}, the free distance growth rates shown in Fig.~\ref{fig:dfreegrowth} are normalized by the \emph{encoding constraint length}. See \cite{mpc13} for further details.} We observe that the convolutional free distance growth rates are significantly larger than the corresponding block minimum distance growth rates for each ensemble. This general technique can be used to find the free distance growth rate of any regular or irregular periodically time-varying protograph-based SC-LDPC-CC ensemble.

The usefulness of the above result is twofold: on the one hand, the fact that the minimum distance growth rates of SC-LDPC-BC ensembles scale to a constant allows us to approximate the growth rates for large $L$ (as noted earlier in Sections \ref{sec:weight} and \ref{sec:bec}), where otherwise it would not be computationally feasible to do so; on the other hand, for large $L$, the free distance is arguably a more appropriate indicator of the strength of convolutional-like SC-LDPC-BC ensembles. In particular, when using convolutional decoding strategies, such as the sliding window decoder discussed above, it is intuitively clear that increasing $L$ will not have a negative effect on performance. In this regard, it is natural that an appropriate distance measure for SC-LDPC-BC ensembles should be independent of $L$, like the free distance growth rate $\delta_\mathrm{free}$ of the associated SC-LDPC-CC ensemble, rather than decaying with $L$, like the minimum distance growth rate $\delta_\mathrm{min}^{(L)}$, which tends to zero as $L \rightarrow \infty$. Numerous empirical studies and simulation results (see, \emph{e.g.}, \cite{lscz10,tss+04,psvc11}) indeed have shown that the performance of SC-LDPC-BCs does not suffer as $L$ is increased, indicating that a distance measure independent of $L$, such as $\delta_\mathrm{free}$, is a more appropriate measure of decoding performance than $\delta_\mathrm{min}$.


\section{Concluding Remarks}\label{sec:conc}

In this paper, we have considered protograph-based spatially coupled LDPC codes. By coupling together a series of $L$ disjoint, or uncoupled, block protographs into a single coupled chain by means of an edge spreading operation, we introduce memory into the code design and obtain the graph of a SC-LDPC-BC ensemble. By varying $L$, we obtain a flexible family of code ensembles with varying rates and code properties that can share the same encoding and decoding architecture for arbitrary $L$. For the $\mathcal{C}(J,K,L)$ ensembles, despite being almost regular, we demonstrated that the resulting codes combine the best features of optimized irregular and regular codes in one design: capacity approaching iterative BP decoding thresholds \emph{and} linear growth of minimum distance with block length. In particular, we saw that, for sufficiently large $L$, the BP thresholds on both the BEC and AWGNC \emph{saturate} to a value significantly larger than the BP decoding threshold and numerically indistinguishable from the MAP decoding threshold of the underlying LDPC-BC ensemble. 
Since all variable nodes have degree greater than two, asymptotically the error probability converges at least doubly exponentially with decoding iterations, and we obtain sequences of asymptotically good LDPC codes with fast convergence rates and BP thresholds close to the Shannon limit. The gap to the Shannon limit  decreases as the density of the graph increases, opening up a new way to construct capacity achieving codes on MBS channels with low-complexity BP decoding.

The key to the excellent threshold performance of SC-LDPC-BC ensembles is a slight structured irregularity introduced to the graph at the boundaries. As $L$ increases, we obtain a family of codes with increasing design rates and a trade-off between capacity approaching iterative decoding thresholds and declining minimum distance growth rates. However, we saw that the growth rates, while declining with $L$, converge to a bound on the free distance growth rate of the closely related SC-LDPC-CC ensemble, which is independent of $L$ and significantly larger than the minimum distance growth rate of the underlying LDPC-BC ensemble, indicating that, particularly in conjunction with convolutional decoding strategies such as a sliding window decoder, an appropriate distance measure for SC-LDPC-BC ensembles should also be independent of $L$. Finally, we showed that the threshold saturation effect obtained by spatial coupling a sequence of disjoint graphs is a general phenomenon and can be applied to both regular and irregular LDPC-BC ensembles. Moreover, carefully designing the edge spreading, increasing the density of the component  graphs, and coupling optimized irregular graphs can further improve performance in terms of both asymptotic minimum distance growth rate and iterative BP decoding threshold.



%





\ifCLASSOPTIONcaptionsoff
  \newpage
\fi



%
\bibliographystyle{IEEEtran}

%








\end{document}